\documentclass{article}
\usepackage{graphicx}
\usepackage[centertags]{amsmath}
\usepackage{amsfonts}
\usepackage{amssymb}
\usepackage{amsthm}

\title{Quantized excitations of internal affine modes and their influence on Raman spectra}

\author{J. J. S\l awianowski, V. Kovalchuk,\\ 
B. Go\l ubowska, A. Martens and E. E. Ro\.zko\\
Institute of Fundamental Technological Research,\\
Polish Academy of Sciences,\\
$5^{\rm B}$, Pawi\'{n}skiego str., 02-106 Warsaw, Poland\\
e-mails: jslawian@ippt.gov.pl, vkoval@ippt.gov.pl,\\ 
bgolub@ippt.gov.pl, amartens@ippt.gov.pl, erozko@ippt.gov.pl}

\begin{document}

\maketitle
\begin{abstract}
Discussed is the structure of classical and quantum excitations of internal degrees of freedom of multiparticle objects like molecules, fullerens, atomic nuclei, etc. Basing on some invariance properties under the action of isometric and affine transformations we reviewed some new models of the mutual interaction between rotational and deformative degrees of freedom. Our methodology and some results may be useful in the theory of Raman scattering and nuclear radiation.
\end{abstract}
  
\section*{Introduction}

Roughly speaking, in what follows we are dealing with some ideas concerning applications of affine symmetry in fundamental physics. The very idea is not new. Let us mention, e.g., some papers by Hehl, Ne'eman, \v{S}ija\v{c}ki and others (see \cite{D.59,D.52,D.53,D.60,D.19,D.20,D.62,D.61,D.33} and references therein).

Everybody knows from ancient times that geometry of the physical world is organised in a hierarchical way and consists of some modules. The most elementary of them is one based on the ideas of Thales of Miletus. The crucial concepts and ideas there were just the parallelism and proportion of segments on the same straight line or, to be more precise, on parallel straight lines. Much of our geometry may be developed in this way. And further on, the "brutal" metrical concepts of distances and angles, or equivalently --- scalar products, did appear. Mathematically speaking, they have to do with bilinear forms on the space of translations in the physical space. No doubt, the metrical concepts are very important also on the fundamental level. Indeed the very pseudo-Euclidean metric structure of the Minkowskian four-dimensional space-time or the Euclidean three-dimensional space are, so to speak, a vacuum-state manifestation ("ground state" in a sense) of the very important field, i.e., just the gravitational field, the force known to humanity from ancient times. Nevertheless, if the Thales-Euclid hierarchy of concepts is seriously treated, it seems very natural to ask what would be the corresponding hierarchy of physical concepts and theories. Then one can ask, even if motivated by a purely academic interest, what would be a hypothetical affine physics and what natural schemes of the metrical (isometry-based) breaking of that Thales-preestablished affine harmony  seem to be physically and experimentally justified. And why at all? Is this some additional interaction or something like the spontaneous symmetry breaking?

The idea of using ${\rm GL\left(3,\mathbb{R}\right)}$, ${\rm SL\left(3,\mathbb{R}\right)}$ or other non-compact groups, e.g., symplectic and pseudo-orthogonal ones, is not new in fundamental physics, let us refer again to Hehl, Ne'eman and \v{S}ija\v{c}ki et. al. \cite{D.59,D.52,D.53,D.60,D.19,D.20,D.62,D.61}. The pseudo-orthogonal group ${\rm O\left(2,4\right)}$ is related in a known way to the theory of hydrogen atom. In nuclear physics the group ${\rm GL\left(3,\mathbb{R}\right)}$ (or rather its subgroup ${\rm SL\left(3,\mathbb{R}\right)}$; nuclear matter seems to be as incompressible as usual macroscopic fluids) has been used for a long time in the collective droplet model of nuclei \cite{BohrMot_69,BohrMot_75,D.46,Dav_58,EisGr_70,D.40,D.41,D.42,D.43,D.44,D.45,D.39,WildTang_77}. Some of the mentioned non-compact groups were thought on as configuration spaces. They were also used as so-called non-invariance groups, spectrum-generating groups (or algebras), e.g., Hamiltonian was one of the generators, and the spectrum might have been found in a rather algebraic way, on the basis of commutation relations alone, etc. (see, e.g., \cite{D.18,D.5,LanLif,D.17} and references therein).

In a sense the classical model of affine degrees of freedom has been used for ages in the theory of the Earth's shape (Riemann, Dedekind, Maclaurin; it seems that in a sense even Newton himself; see, e.g., \cite{D.35,D.4,D.38} and references therein). In XX-th century the interest in this topic revived, as mentioned, in the connection with nuclear and molecular dynamics \cite{Marg_69,Slat_68,Slat_74,all_04,all_05}, astrophysics \cite{D.35} and the theory of continua with microstructure \cite{D.23,D.6,D.25,D.24,all_04,all_05,D.2}. The famous concept of ellipsoidal figures of equilibrium became fashionable. On the purely mathematical level this effort was accompanied by new methods based on Poisson geometry and the dynamics on co-adjoint orbits of Lie groups. The subject has been intensively studied both on the classical and quantum levels \cite{Abr_78,Arn_78,D.11,D.12,D.16,D.14,D.15,D.13,all_04,all_05}.

As mentioned, we concentrate on the linear group as one which rules degrees of freedom of internal/relative motion. So, the only admissible modes of motion are spatial translations, rotations and homogeneous deformations. In a sense, our topic belongs to the dynamics on Lie groups as developed by Arnold, Marsden, Hermann and others \cite{Abr_78,Arn_78,D.36,D.37}, in particular, something like "affine Euler equations" appears \cite{all_04,all_05}. But it is so "in a sense" only. Degrees of freedom are ruled by Lie groups ${\rm GL\left(3,\mathbb{R}\right)}$, ${\rm GAf\left(3,\mathbb{R}\right)}$, but it is not so with Lagrangians, metrics on the configuration space and equations of motion. But the very taste of systems with group-theoretic degrees of freedom consists in considering left or right (or both, i.e., two-sides) invariant geodetic models. It happens quite often that some explicit solutions may be analytically obtained and expressed in terms of exponential mappings on Lie algebras or in terms of some well-investigated special functions on groups \cite{D.11,D.12,all_04,all_05,D.51,D.50,D.34}.

But for all applications and models mentioned above the affine or linear groups do not preserve Lagrangians, Hamiltonians and equations of motion. The corresponding metrics on the configuration space have much weaker, at most isometric symmetry (Euclidean motions, preserving the Euclidean metrics of the physical space). So, and this is really disappointing, there is no profit of using group-theoretic degrees of freedom. Hamiltonian generators of the total affine group in general fail to be constants of motion or good quantities to be used in balance laws when external influences are taken into account. And geodetic models, i.e., ones without forces and potentials, are not useful, because in non-compact spaces they, as a rule, predict the unbounded, "escaping to infinity", i.e., "dissociative", behaviour and no bounded motion, i.e., no oscillatory "elastic" phenomena. Similarly, on the quantum level one obtains purely continuous spectrum, no bound states.

Let us compare all that with the beautiful structure of left-invariant geodetic models on the rotation group, i.e., with the translations-free rigid body (pure rotations) and with the right-invariant geodetic models on the group of volume-preserving diffeomorphisms, i.e., with the Arnold description of incompressible ideal fluids \cite{Abr_78,Arn_78,D.22,D.28,D.29,D.30,D.31}. And we try to repeat this reasoning for affine bodies. Quite independently of any perspective of physical applications, it would be interesting at least from the purely academic point of view of analytical and quantum mechanics to investigate the theory of systems with the kinetic energy (Riemann structure of the configuration space) just exactly invariant under spatial (Euler coordinates) or material (Lagrange coordinates) affine group. Let us also mention that the model right-invariant under ${\rm SL}\left(n,\mathbb{R}\right)$ (under unimodular and orientation-preserving transformations of Lagrange variables) may be considered as a very drastic discretization of the Arnold description of ideal incompressible fluids. Discretization is drastic, because it reduces the continuous system with infinite number of degrees of freedom to one with a finite number of them, namely $n(n+1)$ in the $n$-dimensional "physical" space, i.e., $12$ in the physical case when $n=3$; among them $3$ translational and $9$ internal (relative motion) degrees of freedom (respectively $n$ and $n^2$ in the academic $n$-dimensional world). Applications, e.g., to "nuclear fluids" seem quite possible. Affine invariance in Euler variables is a bit more doubtful and academic, nevertheless, there is some logical necessity which tells us that such models must be formulated (that is done below) and their consequences must be carefully derived and compared with the experimental data (the last thing yet to be done). In any case, quite independently on the academic idea of affinely-invariant models based on the Thales part of geometric axioms, there are some physical arguments which encourage us to undertake this effort. There are known situations where in condensed matter, especially in strongly condensed matter, the effective kinetic energy is not based on the "true" metric tensor, but on the tensor depending somehow on physical parameters. The most important example is the effective mass tensor of electrons in solid state physics \cite{Kit_71}. The point is that the inertia of electrons becomes then less dependent on the spatial metric tensor and more sensitive to surrounding physical factors influencing their motion \cite{Kit_71}. Some phenomena like those are expected also in dynamics of defects in solids or in the theory of a bit exotic objects like bubbles and voids in continua \cite{D.23}, e.g., in fluids. The more so, such exotic fluids like nuclear matter may perhaps behave in a way compatible with predictions of affinely left-invariant models (ones affinely invariant in the physical space). In any case some interesting things are expected when, e.g., the effective mass tensor is proportional to the Cauchy deformation tensor. Geometrically this is the simplest model affinely invariant in the physical space.

It must be strongly stressed, and the above introductory remarks prove this, that the models analysed below, in spite of their using of ${\rm GL}\left(3,\mathbb{R}\right)$ or ${\rm GAf}\left(3,\mathbb{R}\right)$  and their subgroups, have little, if anything, to do with otherwise very respectable and physically useful historical models discussed classically by Riemann, Dedekind, Dirichlet, Chandrasekhar, Bogoyavlenski (\cite{D.35,all_04,all_05} and references therein), and on the quantum microphysical level by Weaver, Cusson, Biedenharn, Rosenstell, Rove and many others \cite{BohrMot_69,BohrMot_75,D.46,Dav_58,EisGr_70,D.40,D.41,D.42,D.43,D.44,D.45,D.39,WildTang_77}. Their dynamics are completely different. Because of this there are also some differences in analytical methods. When dealing with matrix groups, calculations are usually based on various decompositions of groups. For our purposes the most convenient ones are polar and, first of all, singular value decompositions; for latter one we often use the suggestive term "two-polar decomposition". Other well-known multiplicative splittings like the Gauss or Iwasawa decompositions do not seem to be effective within the framework of our models.

The peculiarity of our model is that, according to the Noether theorem, affine momenta, i.e., generators of affine group and their functions, are quantum constants of motion. More precisely, it is so for generators of the volume-preserving subgroup. In particular, if translational motion is neglected, the Casimirs of ${\rm SL}(n,\mathbb{R})$ are constants of motion \cite{all_04,all_05}.

Another very interesting and important feature of affinely-invariant kinetic energies is that they may be used as Lagrangians describing the dynamics of large isochoric elastic vibrations coupled with rotational motion. No potential term is necessary in such models and all advantages of invariant geodetic dynamical systems on Lie groups or their homogeneous spaces may be used \cite{Abr_78,Arn_78,D.11,D.12,D.16,D.14,D.15,D.13,all_04,all_05}. The idea of encoding the dynamics in an appropriately chosen metric tensor of the configuration space resembles the Maupertuis variational principle, however, in our model the mentioned metrics are very natural and follow from some first principles based on geometric ideas. If the object is compressible, then dilatations must be stabilized by some potential term depending on the dilatational ratio, i.e., on ${\rm det}\varphi$ in the formula (\ref{eq1}) below. It is natural to assume something like the potential well or other model rapidly growing when ${\rm det}\varphi$ deviates from unity \cite{D.28}.

It must be stressed that our constituent point particles are spinless. The term "spin" in the classical and quantum parts of our treatment refers to the orbital angular momentum with respect to the centre of mass. When trying to apply the model in nuclear physics one must modify it by introducing in addition and carefully taking into account the "true" spin of constituents, i.e., nucleons. Without such corrections our model may describe only some aspects of the nuclear dynamics. As a matter of fact, the same concerns molecular dynamics, although spin effects are then less essential than in nuclear phenomena.

Incidentally, there are some subtle problems concerning the concept of spin, its affine aspects, and the relationship between spin and orbital angular momentum. And some ideas, perhaps rather speculative and hypothetical, may be formulated. More precisely, we mean here some speculations about the possibility of "explaining" spin as the "orbital" angular momentum of internal motion of extended or composed objects. In standard theories spin is a primary characteristic of elementary particles, described formally in terms of irreducible unitary representations of the group ${\rm SU}(2)$ (${\rm Spin}(n)$ in $n$ dimensions) acting on multicomponent wave functions or fields subject to the quantization procedure. The use of the covering group of ${\rm SO}(3,\mathbb{R})$ (${\rm SO}(n,\mathbb{R})$ in $n$ dimensions) results in appearing of half-integer spin, predicted on the purely experimental, spectroscopic basis by Goudsmit and Uhlenbeck \cite{D.21}. This is angular momentum as an irreducible concept; angular momentum without rotation of anything. There were attempts of formulating affine fundamental physics based on irreducible unitary representations of $\overline{{\rm GL}(3,\mathbb{R})}$, the covering group of ${\rm GL}(3,\mathbb{R})$ ($\overline{{\rm GL}(n,\mathbb{R})}$ in $n$ dimensions) \cite{D.19,D.20}. Of course, those representations are infinite-dimensional; the problem is additionally complicated by the fact that $\overline{{\rm GL}(n,\mathbb{R})}$ is not a linear group (does not possess faithful realizations in terms of finite matrices). When restricted to ${\rm SU}(2)$ (${\rm Spin}(n)$), those representations become reducible and split into direct sums of infinity of irreducible representations found by Wigner and Bargmann \cite{D.52,D.53,D.19,D.20}; obviously, they are finite-dimensional because ${\rm SU}(2)$ (${\rm Spin}(n)$) is compact. "Deformative" generators of $\overline{{\rm GL}(3,\mathbb{R})}$ ($\overline{{\rm GL}(n,\mathbb{R})}$) mix those representations and predict the existence of some excited "trajectories". And of course all characteristics of energy levels are primary and implied by the group itself. When one deals with elementary level, they are not interpreted as following from the rotation and deformation of something in the physical space. The natural question appears as to the possibility of such an interpretation. There were some attempts, among other ones those by Barut, R\c{a}czka and coworkers, of interpreting elementary particles as quantized rotating or rotating-and-deforming objects \cite{D.10,D.9}. Spin quantum numbers would not be then something primary, but rather some aspects of quantized motion in the configuration space of internal degrees of freedom. But how to explain then the half-integer spin? There is some natural hypothesis based on two-valued wave functions. Incidentally, the idea was suggested long ago by Pauli and Reiss \cite{Pauli,Reiss} who claimed that the demand of one-valuedness imposed on wave functions in standard quantum mechanics might be perhaps weakened and sometimes under certain conditions rejected (cf. also \cite{D.47}). This is the case in quantum mechanics of rigid bodies or homogeneously deformable rigid bodies. The classical configuration spaces ${\rm SO}(3,\mathbb{R})$, ${\rm GL}(3,\mathbb{R})$ (${\rm SO}(n,\mathbb{R})$, ${\rm GL}(n,\mathbb{R})$ for $n\geq 3$) are doubly connected and it is quite natural to admit wave functions defined not on them but rather on their covering groups \cite{D.10,D.9,Pauli,Reiss}. This idea has to do with the projective representations of groups. The only point is that to maintain statistical interpretation in ${\rm SO}(3,\mathbb{R})$ (${\rm SO}(n,\mathbb{R})$) or ${\rm GL}(3,\mathbb{R})$ (${\rm GL}(n,\mathbb{R})$), one must assume some kind of superselection. Namely, the kernels of projections from the coverings onto original groups are two-element subgroups. There are two natural subspaces of wave functions: ones which do not distinguish the elements of the kernel subgroup and ones differing in sign there. Functions from two different subspaces cannot be superposed. But it is not clear if statistical interpretation must hold in original groups or in their coverings; in the latter case there is no restriction for superpositions. If proceeding along such lines, one obtains half-integer values of the angular momentum of quantized internal motion. It is interesting that, in a sense, something like this appeared also in our model of quantum torsional oscillator, even without working in the covering groups \cite{D.56}. 

\section{Degrees of freedom, kinematical and canonical variables, symmetries}

We shall consider a system of material points, let us say physically a molecule or perhaps even the atomic nuclei, subject to translational motion, rigid rotations and homogeneous deformations. By this we mean that the current positions of constituent particles in the physical space, analytically speaking its Euler coordinates, are related to the reference Lagrange coordinates $a^{K}$ by the following formula:
\begin{equation}\label{eq1}
y^{i}=x^{i}+\varphi^{i}{}_{K}a^{K},
\end{equation}
where $x^{i}$ are coordinates of the spatial position of a distinguished point of the body, usually the centre of mass, and $\varphi^{i}{}_{K}$ refer to the relative/internal motion. Generalized coordinates are given by the system $\left(x^{i},\varphi^{i}{}_{K}\right)$; just these quantities are considered here as the functions of time. Generalized velocities are given then by the time derivatives $\dot{x}^{i}$, $\dot{\varphi}^{i}{}_{K}$, i.e., respectively the translational velocity and the system of internal ones. Sometimes it is convenient to use the quantities
\begin{equation}\label{eq2}
\Omega^{i}{}_{j}=\dot{\varphi}^{i}{}_{A}\varphi^{-1A}{}_{j},\qquad 
\widehat{\Omega}^{A}{}_{B}=\varphi^{-1A}{}_{i}\dot{\varphi}^{i}{}_{B}=
\varphi^{-1A}{}_{i}\Omega^{i}{}_{j}\varphi^{j}{}_{B},
\end{equation}
so-called affine velocities respectively in spatial and co-moving representations. If $g_{ij}$, $\eta_{AB}$ are metric coefficients respectively in the physical and material spaces (Euler and Lagrange metrics), then the corresponding doubled skew-symmetric parts, 
\begin{equation}\label{eq3}
\Omega^{i}{}_{j}-\Omega_{j}{}^{i}=\Omega^{i}{}_{j}-
g^{ik}g_{jl}\Omega^{l}{}_{k},\quad \widehat{\Omega}^{A}{}_{B}-\widehat{\Omega}_{B}{}^{A}=\widehat{\Omega}^{A}{}_{B}-
\eta^{AC}\eta_{BD}\widehat{\Omega}^{D}{}_{C},
\end{equation}
may be interpreted as angular velocities. Obviously, the upper case indices in $g$ and $\eta$ refer to contravariant inverses of $g_{ij}$ and $\eta_{AB}$, i.e.,
\begin{equation}\label{eq4}
g^{ik}g_{kj}=\delta^{i}{}_{j},\qquad \eta^{AC}\eta_{CB}=\delta^{A}{}_{B}.
\end{equation}
Usually, although not always, one uses orthonormal coordinates in which
\begin{equation}\label{eq5}
g_{ik}=_{\ast}\delta_{ik},\qquad \eta_{AB}=_{\ast}\delta_{AB}.
\end{equation}
{\bf Remark:} unlike in (\ref{eq2}), the skew-symmetric parts (\ref{eq3}) in general are not $\varphi$-related,
\begin{equation}\label{eq6}
\Omega^{i}{}_{j}-\Omega_{j}{}^{i}\neq
\varphi^{i}{}_{A}\left(
\widehat{\Omega}^{A}{}_{B}-\widehat{\Omega}_{B}{}^{A}
\right)\varphi^{-1B}{}_{j}.
\end{equation}
The equality holds if and only if $\varphi$ is an isometry, i.e.,
\begin{equation}\label{eq7}
\eta_{AB}=g_{ij}\varphi^{i}{}_{A}\varphi^{j}{}_{B}.
\end{equation}
Then $\Omega^{i}{}_{j}$, $\widehat{\Omega}^{A}{}_{B}$ are respectively $g$- and $\eta$-skew-symmetric and coincide with the angular velocity in spatial and co-moving representations.

An important point is that the quantities $\Omega^{i}{}_{j}$, $\widehat{\Omega}^{A}{}_{B}$ are non-holonomic velocities or quassivelocities in the sense of Boltzmann. By this we mean that there are no generalized coordinates $x^{i}{}_{j}$ or $y^{A}{}_{B}$ such that $\Omega^{i}{}_{j}$, $\widehat{\Omega}^{A}{}_{B}$ would be their time derivatives. Quasivelocities are linear functions of generalized velocities, however, with configuration-dependent coefficients.

In certain formulae it is convenient to use the co-moving components of the translational velocity,
\begin{equation}\label{eq8}
\widehat{v}^{A}=\varphi^{-1A}{}_{i}v^{i}=\varphi^{-1A}{}_{i}\frac{dx^{i}}{dt}.
\end{equation}

The Green and Cauchy deformation tensors are respectively given by \cite{D.7,D.8}
\begin{equation}\label{eq9}
G_{AB}=g_{ij}\varphi^{i}{}_{A}\varphi^{j}{}_{B},\qquad C_{ij}=\eta_{AB}\varphi^{-1A}{}_{i}\varphi^{-1B}{}_{j},
\end{equation}
and their inverses are as follows: 
\begin{equation}\label{eq10}
G^{-1AB}=\varphi^{-1A}{}_{i}\varphi^{-1B}{}_{j}g^{ij},\qquad C^{-1ij}=\varphi^{i}{}_{A}\varphi^{j}{}_{B}\eta^{AB}.
\end{equation}
{\bf Warning:} to avoid mistakes it is better not to omit the inverse label at $C^{-1}$, $G^{-1}$. Otherwise some confusions would be possible with the $g$- and $\eta$-shifts of indices. And again it is typical that those are different things and except some special situations the following inequalities hold:
\begin{equation}\label{eq12}
C^{-1ij}\neq g^{ik}g^{jl}C_{kl},\qquad G^{-1AB}\neq \eta^{AC}\eta^{BD}G_{CD}.
\end{equation}
Having at disposal two pairs of twice covariant tensors $(g,C)$, $(\eta,G)$, we can construct two mixed tensors
\begin{equation}\label{eq13}
\widehat{C}^{i}{}_{j}=g^{ik}C_{kj},\qquad \widehat{G}^{A}{}_{B}=\eta^{AC}G_{CB}
\end{equation}
and the corresponding basic deformation invariants \cite{D.7,D.8}
\begin{equation}\label{eq14}
I_{p}={\rm Tr}\left(\widehat{G}^{p}\right)={\rm Tr}\left(\widehat{C}^{-1p}\right),\qquad p=1,\ldots,n.
\end{equation}
The quantity $I_{p}$ for any other value of $p$ is a function of the above ones. This is a consequence of the Cayley-Hamilton theorem. For non-deformed configurations the quantities $G$, $C$ coincide  respectively with $\eta$, $g$. Sometimes it is more convenient to use measures of deformations vanishing at the non-deformed configurations, so-called Lagrange and Euler deformation tensors \cite{D.7,D.8}:
\begin{equation}\label{eq15}
E=\frac{1}{2}\left(G-\eta\right),\qquad e=\frac{1}{2}\left(g-C\right).
\end{equation}
Constructing from them the mixed tensors $\widehat{E}$, $\widehat{e}$, i.e.,
\begin{equation}\label{eq16}
\widehat{E}^{A}{}_{B}=\eta^{AC}E_{CB},\qquad \widehat{e}^{i}{}_{j}=g^{ik}e_{kj},
\end{equation}
we can obtain new versions of basic deformation invariants
\begin{equation}\label{eq17}
{\rm Tr}\left(\widehat{E}^{p}\right),\qquad {\rm Tr}\left(\widehat{e}^{p}\right),\qquad p=1,\ldots,n.
\end{equation}
They have the advantage of vanishing when there is no deformation; but of course there is nothing essentially new in them. They are some functions of (\ref{eq14}).

Transformation groups and symmetries are very important for our analysis. Linear spatial and material transformations act on internal degrees of freedom as follows:
\begin{equation}\label{eq18}
\varphi\mapsto\left(L_{A}\circ R_{B}\right)(\varphi)=
\left(R_{B}\circ L_{A}\right)(\varphi)=A\varphi B,
\end{equation}
analytically
\begin{equation}\label{eq19}
\left[\varphi^{i}{}_{K}\right]\mapsto\left[
A^{i}{}_{j}\varphi^{j}{}_{L}B^{L}{}_{K}\right].
\end{equation}
If $A$, $B$ are isometries, i.e.,
\begin{equation}\label{eq20}
g_{ij}A^{i}{}_{m}A^{j}{}_{n}=g_{mn},\qquad \eta_{KL}B^{K}{}_{M}B^{L}{}_{N}=\eta_{MN},
\end{equation}
then obviously $I_{p}$ are preserved by (\ref{eq18}):
\begin{equation}\label{eq21}
I_{p}\left(A\varphi B\right)=I_{p}(\varphi);
\end{equation}
this is just the reason why they are called deformation invariants.

Affine velocities $\Omega$, $\widehat{\Omega}$ suffer the following transformation rule under (\ref{eq18}):
\begin{equation}\label{eq19a}
\Omega\mapsto A\Omega A^{-1},\qquad \widehat{\Omega}\mapsto B^{-1}\widehat{\Omega}B.
\end{equation}
Similarly
\begin{equation}\label{eq11a}
\left[v^{i}\right]\mapsto\left[A^{i}{}_{j}v^{j}\right],\qquad \left[\widehat{v}^{K}\right]\mapsto\left[B^{-1K}{}_{M}\widehat{v}^{M}\right].
\end{equation}
Euler velocity field in the physical space may be expressed in the following way through the quantity $\Omega^{i}{}_{j}$,
\begin{equation}\label{eq12a}
v^{i}(y)=v^{i}+\Omega^{i}{}_{j}\left(y^{j}-x^{j}\right),
\end{equation}
i.e., $\Omega^{i}{}_{j}$ is the gradient of $y\rightarrow v(y)$ in affine motion. Similarly, $\dot{\varphi}^{i}{}_{A}$ is related to the Lagrange velocity field,
\begin{equation}\label{eq13a}
V^{i}(a)=v^{i}+V^{i}{}_{K}a^{K}=\frac{dx^{i}}{dt}+
\frac{d\varphi^{i}{}_{K}}{dt}a^{K}.
\end{equation}
Having in view analysis of equations of motion and first of all the quantization procedure, we must mention the basic concepts of Hamiltonian formalism. Canonical momenta conjugate to $x^{i}$, $\varphi^{i}{}_{A}$, or more precisely dual to generalized velocities $\dot{x}^{i}$, $\dot{\varphi}^{i}{}_{A}$, will be denoted respectively by $p_{i}$, $P^{A}{}_{i}$. In certain formulae it is convenient to use the co-moving representation of translational motion:
\begin{equation}\label{eq14a}
\widehat{p}_{A}=p_{i}\varphi^{i}{}_{A}.
\end{equation}
Instead of $P^{A}{}_{i}$ it is more convenient to use non-holonomic canonical momenta of internal motion conjugate to non-holonomic velocities $\Omega^{i}{}_{j}$, $\widehat{\Omega}^{A}{}_{B}$. We shall denote them respectively by $\Sigma^{i}{}_{j}$, $\widehat{\Sigma}^{A}{}_{B}$:
\begin{equation}\label{eq15a}
\Sigma^{i}{}_{j}=\varphi^{i}{}_{A}P^{A}{}_{j},\qquad \widehat{\Sigma}^{A}{}_{B}=P^{A}{}_{i}\varphi^{i}{}_{B}=
\varphi^{-1A}{}_{i}\Sigma^{i}{}_{j}\varphi^{j}{}_{B}.
\end{equation}
These quantities are referred to as components of affine spin with respect to the space- and body-fixed axes respectively. They are Hamiltonian generators (momentum mappings) of transformations (\ref{eq18}), (\ref{eq19a}) \cite{Abr_78,Arn_78,Gold_50,D.57}. Left and right acting (in the sense of (\ref{eq19a})) rotation subgroups ${\rm SO}\left(n,g\right)$, ${\rm SO}\left(n,\eta\right)$ are generated by the corresponding skew-symmetric parts,
\begin{eqnarray}
S^{i}{}_{j}&=&\Sigma^{i}{}_{j}-\Sigma_{j}{}^{i}=
\Sigma^{i}{}_{j}-g_{jk}g^{il}\Sigma^{k}{}_{l},\label{eq20a}\\
V^{A}{}_{B}&=&\widehat{\Sigma}^{A}{}_{B}-\widehat{\Sigma}_{B}{}^{A}=
\widehat{\Sigma}^{A}{}_{B}-\eta^{AC}\eta_{BD}\widehat{\Sigma}^{D}{}_{C},
\label{eq21a}
\end{eqnarray}
i.e., by the canonical spin $S$ and vorticity $V$ (in Dyson terms \cite{Dys_68}).

\noindent{\bf Warning} (like in (\ref{eq6})): $V^{A}{}_{B}$ are NOT (!) co-moving components of $S$, i.e.,
\begin{equation}\label{eq22}
V^{A}{}_{B}\neq \varphi^{-1A}{}_{i}S^{i}{}_{j}\varphi^{j}{}_{B}.
\end{equation}

The mentioned duality between non-holonomic canonical momenta and non-holonomic velocities is meant in the obvious sense
\begin{equation}\label{eq23}
\Sigma^{i}{}_{j}\Omega^{j}{}_{i}+p_{i}v^{i}=
\widehat{\Sigma}^{A}{}_{B}\widehat{\Omega}^{B}{}_{A}+
\widehat{p}_{A}\widehat{v}^{A}=
P^{A}{}_{i}V^{i}{}_{A}+p_{i}v^{i}
\end{equation}
for all virtual velocities.

The non-holonomic character of $\Sigma^{i}{}_{j}$ or $\widehat{\Sigma}^{A}{}_{B}$ consists in that their Poisson brackets do not vanish; cf. below the formulae (\ref{eq49a}), (\ref{eq49b}).

In practical calculations, especially ones concerning problems with high dynamical symmetries, one uses special systems of generalized coordinates based on the polar and two-polar (singular value) decompositions. They are motivated physically by the classification of degrees of freedom and distinction between various modes of motion.

Roughly speaking, the polar decomposition consists in representing $\varphi$ as a product of linear isometry (orthogonal matrix) and positively definite symmetric matrix in that or opposite ordering; so, there are two versions of this decomposition.

More precisely, having two pairs of "metrics" $\left(G[\varphi],\eta\right)$, $\left(C[\varphi],g\right)$ for any internal configuration $\varphi$, we can express $G[\varphi]$, $C[\varphi]$ through normalised orthogonal principal axes (in the generic case those axes are unique):
\begin{eqnarray}
G[\varphi]&=&\sum_{a}\lambda_{a}F^{a}[\varphi]\otimes F^{a}[\varphi],\label{eq24a}\\
C[\varphi]&=&\sum_{a}\lambda^{-1}_{a}f^{a}[\varphi]\otimes f^{a}[\varphi],\label{eq24b}
\end{eqnarray}
where $F_{a}[\varphi]$, $f_{a}[\varphi]$ are vectors of orthonormal bases (respectively in $\eta$- and $g$-sense) diagonalizing the deformation tensors $G[\varphi]$, $C[\varphi]$, and $F^{a}[\varphi]$, $f^{a}[\varphi]$ are elements of the corresponding dual bases. 

Let us mention that in this way the $\varphi$-degrees of freedom split into three subsystems: two fictitious rigid bodies (given by principal axes of the Green and Cauchy deformation tensors and deformation invariants, i.e., pure stretchings $\lambda_{a}$). There is exactly one isometry $U[\varphi]$ such that
\begin{equation}\label{eq25}
U[\varphi]F_{a}[\varphi]=f_{a}[\varphi],\qquad a=1,\ldots,n.
\end{equation}
It is clear that
\begin{equation}\label{eq26}
g_{ij}U[\varphi]^{i}{}_{K}U[\varphi]^{j}{}_{L}=\eta_{KL}
\end{equation}
and
\begin{equation}\label{eq27}
\varphi=U[\varphi]A[\varphi],
\end{equation}
where $A[\varphi]$ is $\eta$-symmetric,
\begin{equation}\label{eq28}
\eta_{KM}A[\varphi]^{M}{}_{L}=\eta_{LM}A[\varphi]^{M}{}_{K},
\end{equation}
and positively definite.

As mentioned, one can also write
\begin{equation}\label{eq29}
\varphi=B[\varphi]U[\varphi],\qquad B[\varphi]=U[\varphi]A[\varphi]U[\varphi]^{-1},
\end{equation}
where $B[\varphi]$ is $g$-symmetric,
\begin{equation}\label{eq30}
g_{ik}B[\varphi]^{k}{}_{j}=g_{jk}B[\varphi]^{k}{}_{i},
\end{equation}
and positively definite. It is well known that the polar decomposition in both versions is unique. Analytically this is just the factorization into the product of orthogonal and symmetric-positive matrices. Diagonalizing the symmetric part one obtains the singular value decomposition, i.e., representation of $\varphi$ as a product of orthogonal, positive-diagonal and again orthogonal matrices,
\begin{equation}\label{eq31}
\varphi=LDR^{-1},\qquad L,R\in{\rm SO}\left(n,\mathbb{R}\right),\quad D={\rm Diag}\left(Q^{1},\ldots,Q^{n}\right).
\end{equation}
Using more geometric language, i.e., making a systematic distinction between the material space (Lagrange variables), physical space (Euler variables) and $\mathbb{R}^{n}$ (deformation invariants), we would write that
\begin{equation}\label{eq32}
\varphi^{i}{}_{A}=L^{i}{}_{a}D^{a}{}_{b}R^{-1b}{}_{A},
\end{equation}
where $D:\mathbb{R}^{n}\rightarrow\mathbb{R}^{n}$ is given by
\begin{equation}\label{eq33}
D^{a}{}_{b}=0\quad {\rm if}\quad a\neq b,\qquad D^{\underline{a}}{}_{\underline{a}}=Q^{a}.
\end{equation}
In some formulae it is convenient to use variables $q^{a}$, where
\begin{equation}\label{eq34}
Q^{a}=\exp\left(q^{a}\right).
\end{equation}
The fictitious rigid bodies $L$, $R$, i.e., attitudes of the main axes of inertia have their own angular velocities and canonical spin variables. "Spatial" angular velocities are respectively given by
\begin{equation}\label{eq35}
\chi^{i}{}_{j}=\left(\frac{d}{dt}L^{i}{}_{a}\right)L^{-1a}{}_{j},\qquad \vartheta^{K}{}_{L}=\left(\frac{d}{dt}R^{K}{}_{a}\right)R^{-1a}{}_{L}.
\end{equation}
Much more convenient are their "co-moving" representants:
\begin{equation}\label{eq36}
\widehat{\chi}^{a}{}_{b}=L^{-1a}{}_{i}\left(\frac{d}{dt}L^{i}{}_{b}\right),\qquad \widehat{\vartheta}^{a}{}_{b}=R^{-1a}{}_{K}\left(\frac{d}{dt}R^{K}{}_{b}\right).
\end{equation}
The corresponding canonical spin variables are respectively denoted by
\begin{equation}\label{eq37}
\varrho^{i}{}_{j},\qquad \tau^{K}{}_{L},\qquad \widehat{\varrho}^{a}{}_{b},\qquad \widehat{\tau}^{a}{}_{b}.
\end{equation}
Canonical momenta conjugate to $Q^{a}$, $q^{a}$ are denoted respectively by
\begin{equation}\label{eq38}
P_{a},\qquad p_{a}.
\end{equation}
Let us observe that $Q^{a}$, $q^{a}$, $a=1,\ldots,n$, are alternative measures of deformation invariants. The two-polar splitting is not unique. The invariants $Q^{a}$ or $q^{a}$ are indistinguishable "particles" on $\mathbb{R}$ and may be permuted; those permutations must be accompanied by multiplying the matrices $L$, $R$ on the right by appropriate orthogonal matrices having in each row and column only zeros but once $\pm 1$. More precisely, this finite non-uniqueness is a characteristic feature of the situation when the spectrum of $D$ is non-degenerate, i.e., when all $Q^{i}$-s are pairwise distinct, $Q^{i}\neq Q^{j}$ if $i\neq j$, i.e.,
\begin{equation}\label{eq38a}
\prod_{i\neq j}\left(Q^{i}-Q^{j}\right)\neq 0.
\end{equation}
If degeneracy occurs, i.e., when the above product expression vanishes, then some additional continuous non-uniqueness of the right-multiplying gauge orthogonal matrices also appears. An extreme special case is when all $D^{i}$-s are equal, i.e., when the diagonal matrix $D$ is proportional to identity matrix $I_{n}$. Then $L$, $R$ separately are not well defined and it is only $LR^{-1}$ that is meaningful. The set of orthogonal right-multiplying gauge matrices is identical with the total orthogonal group. And there are of course intermediate situations between the non-degeneracy and the total degeneracy; all they are characterized by some continuous parameters.

It is clear that $\varrho$, $\tau$ coincide with spin and minus vorticity,
\begin{equation}\label{eq39}
\varrho^{i}{}_{j}=S^{i}{}_{j},\qquad \tau^{K}{}_{L}=-V^{K}{}_{L}.
\end{equation}
The pairing between canonical variables and velocities (holonomic and non-holonomic) is given as follows:
\begin{eqnarray}
\left\langle\left(\varrho,\tau,p\right);\left(\chi,\vartheta,\dot{q}\right)
\right\rangle&=&
\left\langle\left(\widehat{\varrho},\widehat{\tau},p\right);
\left(\widehat{\chi},\widehat{\vartheta},\dot{q}\right)\right\rangle
\nonumber\\
=p_{a}\dot{q}^{a}+\frac{1}{2}{\rm Tr}\left(\varrho\chi\right)+\frac{1}{2}{\rm Tr}\left(\tau\vartheta\right)
&=&
p_{a}\dot{q}^{a}+\frac{1}{2}{\rm Tr}\left(\widehat{\varrho}\widehat{\chi}\right)+\frac{1}{2}{\rm Tr}\left(\widehat{\tau}\widehat{\vartheta}\right).\label{eq40}
\end{eqnarray}
Obviously,
\begin{equation}\label{eq41}
p_{a}\dot{q}^{a}=P_{a}\dot{Q}^{a}.
\end{equation}
{\bf Remark:} do not confuse the above $p_{a}$ with translational momenta conjugate to $\dot{x}^{a}$. When there will be a danger of confusion, from now on  canonical momenta conjugate to $\dot{x}^{a}$ will be denoted by $p({\rm tr})_{a}$.

Deformation tensors are expressed by the following formulae:
\begin{eqnarray}
G_{KL}=\eta_{RS}A^{R}{}_{K}A^{S}{}_{L},&\quad& C_{ij}=g_{kl}B^{-1k}{}_{i}B^{-1l}{}_{j}, \label{eq42a}\\
\widehat{G}^{K}{}_{L}=R^{K}{}_{a}D^{a}{}_{c}D^{c}{}_{b}R^{-1b}{}_{L},&\quad& \widehat{C}^{i}{}_{j}=L^{i}{}_{a}D^{-1a}{}_{c}D^{-1c}{}_{b}L^{-1b}{}_{j}. \label{eq42b}
\end{eqnarray}
In purely analytical matrix terms, when orthonormal axes are used and matrices of $\eta$, $g$ coincide with the Kronecker symbol, we can simply write
\begin{equation}\label{eq43}
G=A^{2}=RD^{2}R^{T}=RD^{2}R^{-1},\quad C=B^{-2}=LD^{-2}L^{T}=LD^{-2}L^{-1}.
\end{equation}
One should mention what is the physical meaning of the above decompositions. Roughly speaking, $U[\varphi]$ in (\ref{eq27}), (\ref{eq29}) describes rotational degrees of freedom and $A[\varphi]$, $B[\varphi]$ are two alternative descriptions of homogeneously-deformative modes of motion (equivalent respectively to the Green and Cauchy deformation tensors). But there are two important informations in deformation tensors:
\begin{enumerate}
\item Purely scalar information about the stretching; in $n$-dimensional space there are $n$ independent ones (physically $n=3$, sometimes $2$ or $1$).
\item Information about how this stretching is oriented with respect to the space- or body-fixed reference frames. 
\end{enumerate}
The first information is analytically described by deformation invariants, thus, finally by the diagonal matrix $D$. The "directional" information is equivalent to the knowledge of principal axes of deformation tensors $G$ and $C$ meant in the sense of metric tensors $\eta$ and $g$, i.e., normalised eigenvectors of $\widehat{G}$ and $\widehat{C}$. And this information is contained in objects $L$ and $R$. 

When some origin of the Cartesian coordinate system is fixed in the physical space, by convention the point $y=0$, we can also introduce the orbital and total affine momenta ("hypermomenta" in terminology of Hehl and Ne'eman) \cite{D.52,D.53,D.19,D.20} with respect to that origin. We denote them respectively by
\begin{equation}\label{eq44}
\Lambda^{i}{}_{j}=x^{i}p({\rm tr})_{j},\qquad J^{i}{}_{j}=\Lambda^{i}{}_{j}+\Sigma^{i}{}_{j}=x^{i}p({\rm tr})_{j}+\Sigma^{i}{}_{j}.
\end{equation}
Their doubled skew-symmetric parts are respectively the "orbital" and "total" canonical angular momenta,
\begin{equation}\label{eq45}
L^{i}{}_{j}=x^{i}p({\rm tr})_{j}-x_{j}p({\rm tr})^{i},\quad J^{i}{}_{j}=L^{i}{}_{j}+S^{i}{}_{j}=x^{i}p({\rm tr})_{j}-x_{j}p({\rm tr})^{i}+S^{i}{}_{j}.\
\end{equation}
They are Hamiltonian generators of transformations
\begin{equation}\label{eq46}
{}^{\prime}x^{i}=L^{i}{}_{j}x^{j},\qquad {}^{\prime}\varphi^{i}{}_{K}=L^{i}{}_{j}\varphi^{j}{}_{K},
\end{equation}
where $L$ is orthogonal,
\begin{equation}\label{eq47}
g_{ij}L^{i}{}_{k}L^{j}{}_{m}=g_{km}.
\end{equation}
Similarly, (\ref{eq44}) are Hamiltonian generators for the non-restricted transformations (\ref{eq46}) with the general $L$, without the orthogonal constraints (\ref{eq47}). However, one must always remember that unlike the absolutely defined transformations (\ref{eq18}), (\ref{eq46}) are always related to some fixed origin in the physical space and so are the quantities (\ref{eq45}).

To finish with this kinematical introduction, let us quote the basic Poisson-bracket relations finally following from
\begin{eqnarray}
&&\left\{x^{i},p({\rm tr})_{j}\right\}=\delta^{i}{}_{j},\qquad 
\left\{\varphi^{i}{}_{A},P^{B}{}_{j}\right\}=\delta^{B}{}_{A}\delta^{i}{}_{j},\label{eq48a}\\
&&\left\{x^{i},P^{B}{}_{j}\right\}=0,\qquad 
\left\{x^{i},\varphi^{j}{}_{A}\right\}=0,\qquad \left\{\varphi^{i}{}_{A},p({\rm tr})_{j}\right\}=0.
\label{eq48b}
\end{eqnarray}
The most important of them read
\begin{eqnarray}
&&\left\{\Sigma^{i}{}_{j},\Sigma^{k}{}_{l}\right\}=
\delta^{i}{}_{l}\Sigma^{k}{}_{j}-\delta^{k}{}_{j}\Sigma^{i}{}_{l},\quad 
\left\{\widehat{\Sigma}^{A}{}_{B},\widehat{p}({\rm tr})_{C}\right\}=
\delta^{A}{}_{C}\widehat{p}({\rm tr})_{B},\label{eq49a}\\
&&\left\{\widehat{\Sigma}^{A}{}_{B},\widehat{\Sigma}^{C}{}_{D}\right\}=
\delta^{C}{}_{B}\widehat{\Sigma}^{A}{}_{D}-
\delta^{A}{}_{D}\widehat{\Sigma}^{C}{}_{B},\qquad
\left\{\Sigma^{i}{}_{j},\widehat{\Sigma}^{A}{}_{B}\right\}=0,
\label{eq49b}\\
&&\left\{J^{i}{}_{j},p({\rm tr})_{k}\right\}=\left\{\Lambda^{i}{}_{j},p({\rm tr})_{k}\right\}=
\delta^{i}{}_{k}p({\rm tr})_{j}.\label{eq49c}
\end{eqnarray}
If some function $F$ depends only on the configuration variables $\varphi$, but not on the generalized momenta, then
\begin{equation}\label{eq50}
\left\{F,\Sigma^{i}{}_{j}\right\}=\varphi^{i}{}_{A}\frac{\partial F}{\partial \varphi^{j}{}_{A}},\quad \left\{F,\Lambda^{i}{}_{j}\right\}=x^{i}\frac{\partial F}{\partial x^{j}},\quad \left\{F,\widehat{\Sigma}^{A}{}_{B}\right\}=\varphi^{i}{}_{B}\frac{\partial F}{\partial \varphi^{i}{}_{A}}.
\end{equation}
Non-quoted Poisson brackets do vanish.

The system of Poisson brackets is very helpful when deriving equations of motion in the balance Hamiltonian form,
\begin{equation}\label{eq51}
\frac{dF}{dt}=\{F,H\},
\end{equation}
where $F$ runs over some complete family of Jacobi-independent functions on the phase space.

\section{Inertial properties and canonical symmetries}

The summation of kinetic energies of constituents of our "molecule" results in the following expression:
\begin{equation}\label{eq52}
T=T_{\rm tr}+T_{\rm int}=\frac{m}{2}g_{ij}\frac{dx^{i}}{dt}\frac{dx^{j}}{dt}+ \frac{1}{2}g_{ij}\frac{d\varphi^{i}{}_{A}}{dt}\frac{d\varphi^{j}{}_{B}}{dt}J^{AB},
\end{equation}
where $m$ is the total mass of the body and $J^{AB}$ are, roughly speaking, the co-moving components of inertial tensor or, more precisely, the second-order multipoles of the constant Lagrange distribution of matter in the space of Lagrange coordinates,
\begin{equation}\label{eq53a}
m=\int d\mu(a),\qquad J^{AB}=\int a^{A}a^{B}d\mu(a)=J^{BA}.
\end{equation}
Here $\mu$ is the Lagrange co-moving distribution of mass, automatically constant in time; therefore, also $m$ and $J^{AB}$ are constant in time. 

Obviously, the formula (\ref{eq52}) is valid under the assumption that $x^{i}$ are current coordinates of the centre of mass, i.e., that the origin of Lagrange coordinates, $a^{K}=0$, is chosen in such a way that the dipole moment of $\mu$ vanishes,
\begin{equation}\label{eq52a}
J^{K}=\int a^{K}d\mu(a)=0.
\end{equation}
Otherwise in the kinetic energy expression $T$ some interference, crossing terms bilinear in $\left(dx^{i}/dt,d\varphi^{i}{}_{A}/dt\right)$ would appear. For affine bodies the vanishing of $J^{K}$ is equivalent to the equation
\begin{equation}\label{eq52b}
\int\left(y^{i}-x^{i}\right)d\mu_{(x,\varphi)}(y)=0,
\end{equation}
where $\mu_{(x,\varphi)}$ is the measure obtained from $\mu$ as its transport by the mapping (\ref{eq1}). For non-affine bodies, when (\ref{eq1}) is replaced by a general function $y^{i}\left(a^{K}\right)$, things are more complicated, but there is no place here for their discussion. In any case, the condition $J^{K}=0$ is not then equivalent to taking as $x^{i}$ the coordinates of the current position of the centre of mass in the physical space.

Incidentally, later on (the reasons will become clear after some additional remarks) it is convenient to express (\ref{eq52}) in the following form:
\begin{eqnarray}
T=T_{\rm tr}+T_{\rm int}&=&\frac{m}{2}G_{AB}\widehat{v}^{A}\widehat{v}^{B}+ \frac{1}{2}G_{KL}\widehat{\Omega}^{K}{}_{A}\widehat{\Omega}^{L}{}_{B}J^{AB}
\nonumber\\
&=&\frac{m}{2}g_{ij}v^{i}v^{j}+ \frac{1}{2}g_{ij}\Omega^{i}{}_{k}\Omega^{j}{}_{l}J[\varphi]^{kl},\label{eq53b}
\end{eqnarray}
where $J[\varphi]^{ij}$ are spatial, therefore time-dependent, components of the inertial tensor, i.e.,
\begin{equation}\label{eq54}
J[\varphi]^{kl}=\varphi^{k}{}_{A}\varphi^{l}{}_{B}J^{AB}.
\end{equation}

Incidentally, (\ref{eq54}) may be written in the following form:
\begin{equation}\label{eq54a}
J[\varphi]^{kl}=
\int\left(y^{k}-x^{k}\right)\left(y^{l}-x^{l}\right)d\mu_{(x,\varphi)}.
\end{equation}
This kind of expression is applicable to general bodies and in the case of affine motion it becomes (\ref{eq54}). Although $J[\varphi]$ are components with respect to the inertial frame, apparently more "touchable" than the co-moving (material) one, they are not very useful in dynamical equations. Let us remind that even in elementary problems of rigid body mechanics, we simply must use constant co-moving components $J^{AB}$. Those are either calculated on the basis of our knowledge of the mass distribution in the body or somehow postulated, estimated a priori and testified on the basis of comparison of calculated results with experimental data. 

It is clear that (\ref{eq52}) is invariant under isometries acting in the physical space (parameterized by Euler variables) and under the group ${\rm O}\left(U,J\right)$ preserving the fixed inertial tensor $J$. If $J$ is isotropic in the sense $J^{AB}=I\eta^{AB}$, then we are dealing with the double orthogonal symmetry.

If we use the above-mentioned polar decomposition (\ref{eq27}) and the corresponding "co-moving" angular velocity
\begin{equation}\label{eq55}
\widehat{\omega}=U^{-1}\frac{dU}{dt},
\end{equation}
then the expression for the internal kinetic energy becomes
\begin{equation}\label{eq56}
T_{\rm int}=-\frac{1}{2}{\rm Tr}\left(AJ_{\eta}A\widehat{\omega}^{2}\right)+
{\rm Tr}\left(AJ_{\eta}\frac{dA}{dt}\widehat{\omega}\right)+
\frac{1}{2}{\rm Tr}\left(J_{\eta}\left(\frac{dA}{dt}\right)^{2}\right),
\end{equation}
where $J_{\eta}$ is obtained from $J$ by the $\eta$-lowering of the second index,
\begin{equation}\label{eq57}
J_{\eta}{}^{A}{}_{B}:=J^{AC}\eta_{CB}.
\end{equation}
Obviously, if orthonormal coordinates are used, there is no numerical distinction between matrices of $J_{\eta}$ and $J$. The first term in (\ref{eq56}) is centrifugal one, the second one represents the Coriolis inertial forces, and the third one describes the dynamics of pure deformations. 

If we use the two-polar decomposition (\ref{eq31}) and the body is doubly isotropic,
\begin{equation}\label{eq58}
J^{AB}=I\eta^{AB},
\end{equation}
i.e., in orthonormal coordinates
\begin{equation}\label{eq59}
J^{AB}=_{\ast}I\delta^{AB},
\end{equation}
then the internal kinetic energy (\ref{eq56}) can be rewritten as follows:
\begin{equation}\label{eq60}
T_{\rm int}=-\frac{I}{2}{\rm Tr}\left(D^{2}\widehat{\chi}^{2}\right)
-\frac{I}{2}{\rm Tr}\left(D^{2}\widehat{\vartheta}^{2}\right)+
I{\rm Tr}\left(D\widehat{\chi}D\widehat{\vartheta}\right)+
\frac{I}{2}{\rm Tr}\left(\left(\frac{dD}{dt}\right)^{2}\right),
\end{equation}
where the first two terms are centrifugal ones, the third one describes the Coriolis forces, and the last one is the kinetic energy of pure stretchings. If we perform the Legendre transformation for the Lagrangian
\begin{equation}\label{eq62}
L=T-V(x,\varphi),
\end{equation}
where $T$ is given by (\ref{eq60}), and use the variables
\begin{equation}\label{eq61}
M:=-\widehat{\varrho}-\widehat{\tau},\qquad N:=\widehat{\varrho}-\widehat{\tau},
\end{equation}
then (\ref{eq60}) becomes as follows:
\begin{equation}\label{eq63}
T_{\rm int}=\frac{1}{2I}\sum_{a}P^{2}_{a}
+\frac{1}{8I}\sum_{a\neq b}\frac{\left(M^{a}{}_{b}\right)^{2}}{\left(Q^{a}-Q^{b}\right)^{2}}
+\frac{1}{8I}\sum_{a\neq b}\frac{\left(N^{a}{}_{b}\right)^{2}}{\left(Q^{a}+Q^{b}\right)^{2}}.
\end{equation}

Although everything is in principle clear from the context, it is perhaps convenient and instructive to quote some explicit formulas and statements.

Let $L$ be a Lagrangian of some classical system with generalized coordinates $\xi^{\mu}$ and generalized velocities $\dot{\xi}^{\mu}$. Geometrically speaking, $\left(\xi^{\mu},\dot{\xi}^{\mu}\right)$ are adapted coordinates in the tangent bundle $TQ$, where $Q$ denotes the configuration space, a manifold of generalized "positions". Legendre transformation $\mathcal{L}$ leads from generalized velocities $\dot{\xi}^{\mu}$ to their conjugate momenta $\pi_{\mu}$; in geometric terms, it maps $TQ$ into the cotangent bundle $T^{\ast}Q$ parameterized by $\left(\xi^{\mu},\pi_{\mu}\right)$. Analytically it is given by
\begin{equation}\label{eq76+1}
\left(\xi^{\mu},\dot{\xi}^{\mu}\right)\mapsto
\left(\xi^{\mu},\pi_{\mu}\right)=\left(\xi^{\mu},\frac{\partial L}{\partial \dot{\xi}^{\mu}}\right).
\end{equation}
In "usual" mechanics this mapping is invertible, more precisely, it is a diffeomorphism of $TQ$ onto $T^{\ast}Q$. Inverting it we express generalized velocities in terms of canonical momenta; substituting this expression to the energy function
\begin{equation}\label{eq76+2}
E\left(\xi^{\mu},\dot{\xi}^{\mu}\right)=\dot{\xi}^{\mu}
\frac{\partial L}{\partial \dot{\xi}^{\mu}}-L\left(\xi^{\mu},\dot{\xi}^{\mu}\right),
\end{equation}
one obtains Hamiltonian as a function $H:T^{\ast}Q\rightarrow \mathbb{R}$, analytically, $H(\xi,\pi)$. Later on one can proceed in Hamiltonian terms, following the formula (\ref{eq51}). If Lagrangian has the potential structure (\ref{eq62}) (sometimes called "natural" one), then obviously in (\ref{eq76+1}) $L$ may be replaced by $T$ itself. 

Some comments are necessary concerning the formulas (\ref{eq61}), (\ref{eq63}), the more so that "non-holonomic" quantities $\widehat{\chi}$, $\widehat{\vartheta}$, $\widehat{\varrho}$, $\widehat{\tau}$ are used there. Non-holonomic velocities, or quasivelocities, introduced long ago by Boltzmann and many others, are given by expressions:
\begin{equation}\label{eq76+3}
\omega^{A}=\omega^{A}{}_{\mu}(\xi)\dot{\xi}^{\mu},
\end{equation}
where $\omega^{A}{}_{\mu}(\xi)$ depend on generalized coordinates $\xi$ in an irreducible way, i.e., the Pfaff forms $\omega^{A}{}_{\mu}(\xi)d\xi^{\mu}$ fail to be differentials. In many problems, appropriately chosen $\omega^{A}$ are more convenient than $\dot{\xi}^{\mu}$; typical examples are angular velocity (in space dimension higher than two) and our affine velocity. There is a conjugate concept of non-holonomic momenta, or quasimomenta, $\sigma_{A}$ given by
\begin{equation}\label{eq76+4}
\sigma_{A}=p_{\mu}\sigma^{\mu}{}_{A},
\end{equation}
where the $\omega$- and $\sigma$-matrices are mutually reciprocal,
\begin{equation}\label{eq76+5}
\omega^{A}{}_{\mu}\sigma^{\mu}{}_{B}=\delta^{A}{}_{B},\qquad 
\sigma^{\mu}{}_{A}\omega^{A}{}_{\nu}=\delta^{\mu}{}_{\nu},\qquad
\sigma_{A}\omega^{A}=\pi_{\mu}\dot{\xi}^{\mu}.
\end{equation}
The "non-holonomic" character of $\sigma$-s is that their Poisson brackets do not vanish. Typical examples are angular momentum of rotational motion or our affine momentum ("hyperspin"). And in general, when $Q$ is a Lie group, then it is typical that $\omega$, $\sigma$ are elements of its Lie algebra and co-algebra (dual of a Lie algebra).

When non-holonomic quantities $\omega$, $\sigma$ are used, Legendre transformation (\ref{eq76+2}) may be alternatively expressed as follows:
\begin{equation}\label{eq76+6}
\left(\xi^{\mu},\omega^{A}\right)\mapsto
\left(\xi^{\mu},\sigma_{A}\right)=\left(\xi^{\mu},\frac{\partial L}{\partial \omega^{A}}\right),
\end{equation}
where, obviously, $L$ is to be treated as a function of $\xi$, $\omega$. The energy function $E$ becomes
\begin{equation}\label{eq76+7}
E\left(\xi,\omega\right)=\omega^{A}
\frac{\partial L}{\partial \omega^{A}}-L\left(\xi,\omega\right).
\end{equation}
Inverting Legendre transformation, i.e., expressing $\omega$ through $\sigma$, one obtains Hamiltonian $H\left(\xi,\sigma\right)$.

Just this procedure is meant in formulas (\ref{eq60})--(\ref{eq63}). Configurations $\xi$ are there given by the triples $\left(Q,L,R\right)$ "parameterizing" the object $\varphi$. Non-holono\-mic velocities we use are arrays $\left(\dot{Q}^{a},\widehat{\chi}^{a}{}_{b},\widehat{\vartheta}^{a}{}_{b}\right)$, the kinetic energy is expressed by them in (\ref{eq60}); this is just our specification of the kinetic part of (\ref{eq76+7}), and our non-holonomic conjugate momenta are arrays $\left(P_{a},\widehat{\varrho}^{a}{}_{b},\widehat{\tau}^{a}{}_{b}\right)$. The kinetic part of the Hamiltonian is given by (\ref{eq63}), the total Hamiltonian contains in addition some potential energy term $V\left(x;\varphi\right)=V\left(x;Q,L,R\right)$. Legendre transformation corresponding to (\ref{eq60}) is analytically given by the following expression of $P_{a}$, $\widehat{\varrho}^{a}{}_{b}$, $\widehat{\tau}^{a}{}_{b}$ through $\dot{Q}^{a}$, $\widehat{\chi}^{a}{}_{b}$, $\widehat{\vartheta}^{a}{}_{b}$:
\begin{eqnarray}
P_{a}&=&\frac{\partial T_{\rm int}}{\partial \dot{Q}^{a}}=I\dot{Q}^{a},\label{eq76+8a}\\
\widehat{\varrho}^{a}{}_{b}&=&\frac{\partial T_{\rm int}}{\partial \widehat{\chi}^{b}{}_{a}},\qquad \widehat{\varrho}=\left[\widehat{\varrho}^{a}{}_{b}\right]=
I\left(D^{2}\widehat{\chi}+\widehat{\chi}D^{2}-2D\widehat{\vartheta}D\right),
\label{eq76+8b}\\
\widehat{\tau}^{a}{}_{b}&=&\frac{\partial T_{\rm int}}{\partial \widehat{\vartheta}^{b}{}_{a}},\qquad
\widehat{\tau}=\left[\widehat{\tau}^{a}{}_{b}\right]=
I\left(D^{2}\widehat{\vartheta}+\widehat{\vartheta}D^{2}-2D\widehat{\chi}D\right).
\label{eq76+8c}
\end{eqnarray}
Let us remind that the matrix $D$ is diagonal and $Q^{a}$ are its diagonal elements --- deformation invariants. The last two formulas are a bit "symbolic" because $\widehat{\chi}$, $\widehat{\vartheta}$ are skew-symmetric and their matrix elements are not independent. We mean the following: In expression (\ref{eq60}) we replace $\dot{Q}^{a}$, $\widehat{\chi}$, $\widehat{\vartheta}$ by $\dot{Q}^{a}+\delta\dot{Q}^{a}$, $\widehat{\chi}+\delta\widehat{\chi}$, $\widehat{\vartheta}+\delta\widehat{\vartheta}$ and calculate the corresponding variation of $T_{\rm int}$, i.e., $\delta T_{\rm int}$ up to terms linear in $\delta \dot{Q}^{a}$, $\delta\widehat{\chi}$, $\delta\widehat{\vartheta}$. After some easy calculations we obtain that
\begin{equation}\label{eq76+9}
\delta T_{\rm int}=P_{a}\delta\dot{Q}^{a}+\frac{1}{2}{\rm Tr}\left(\widehat{\varrho}\delta\widehat{\chi}\right)+\frac{1}{2}{\rm Tr}\left(\widehat{\tau}\delta\widehat{\vartheta}\right),
\end{equation}
where the skew-symmetric matrices $\widehat{\varrho}$, $\widehat{\tau}$ are given just in (\ref{eq76+8a})--(\ref{eq76+8c}). Compare this also with (\ref{eq40}), (\ref{eq41}). The space of skew-symmetric matrices is isomorphic with its own dual in the sense of the above trace expressions, and in this sense the last two formulas in (\ref{eq76+8a})--(\ref{eq76+8c}) are meant, because the differential of a function is by definition given by the main part of its variation, i.e., one linear in variation of independent variables. The factors $1/2$ in (\ref{eq40}),  (\ref{eq76+8a})--(\ref{eq76+8c}) are due to the fact that because of the skew-symmetry of $\widehat{\chi}$, $\widehat{\vartheta}$, $\widehat{\varrho}$, $\widehat{\tau}$ every term in the traces in (\ref{eq76+9}) occurs twice. In the physical dimension $n=3$ (but neither in planar problems $n=2$, nor in the "academic" situation $n\geq 3$), the above formulas may be written in a simpler and more familiar form, when $\widehat{\chi}$, $\widehat{\vartheta}$, $\widehat{\varrho}$, $\widehat{\tau}$ are identified with $3$-dimensional pseudovectors (axial vectors); do not confuse them with the usual ("non-pseudo") vectors:
\begin{eqnarray}
\widehat{\chi}^{a}{}_{b}=-\varepsilon^{a}{}_{bc}\widehat{\chi}^{c},&\quad& \widehat{\vartheta}^{a}{}_{b}=-\varepsilon^{a}{}_{bc}\widehat{\vartheta}^{c},
\label{eq76+10a}\\
\widehat{\varrho}^{a}{}_{b}=\varepsilon^{a}{}_{b}{}^{c}
\widehat{\varrho}_{c},&\quad& \widehat{\tau}^{a}{}_{b}=\varepsilon^{a}{}_{b}{}^{c}
\widehat{\tau}_{c}.\label{eq76+10b}
\end{eqnarray}
Obviously, $\varepsilon$ denotes here the totally antisymmetric Ricci symbol and the shift of indices is meant in the trivial sense of the Kronecker-delta metric. Then (\ref{eq76+8a})--(\ref{eq76+9}) become respectively as follows: 
\begin{eqnarray}
P_{a}&=&\frac{\partial T_{\rm int}}{\partial \dot{Q}^{a}}=I\dot{Q}^{a},\label{eq76+11a}\\
\widehat{\varrho}_{a}&=&\frac{\partial T_{\rm int}}{\partial \widehat{\chi}^{a}},\label{eq76+11b}\\
\widehat{\tau}_{a}&=&\frac{\partial T_{\rm int}}{\partial \widehat{\vartheta}^{a}},\label{eq76+11c}\\
\delta T_{\rm int}&=&P_{a}\delta\dot{Q}^{a}+\widehat{\varrho}_{a}\delta\widehat{\chi}^{a}+
\widehat{\tau}_{a}\delta\widehat{\vartheta}^{a}.\label{eq76+11d}
\end{eqnarray}
Solving (\ref{eq76+8a})--(\ref{eq76+8c}) with respect to $\dot{D}$, $\widehat{\chi}$, $\widehat{\vartheta}$ one obtains (\ref{eq63}) with $M$, $N$ expressions defined in (\ref{eq61}).

As seen from the very form of $T_{\rm tr}$ in (\ref{eq52}), the Lagrangian given by $T_{\rm tr}$ alone, without the potential term $V(x,\varphi)$, is non-physical in elasticity and condensed matter theory, because it predicts an unbounded expansion of the body. This is also seen in (\ref{eq63}) where the interaction between deformation invariants is purely repulsive.

And at the same time there is some "aesthetic" drawback of (\ref{eq52}) and the mentioned resulting formulae. Namely, in spite of the affine symmetry of degrees of freedom, the kinetic energy is not affinely invariant in the sense of dynamics. Kinetic energies, i.e., the metric tensors on the configuration space, invariant under the left or right affine translations are respectively given by the following expressions:
\begin{eqnarray}
T=T_{\rm tr}+T_{\rm int}&=&\frac{m}{2}\eta_{AB}\widehat{v}^{A}\widehat{v}^{B}+
\frac{1}{2}\mathcal{L}^{A}{}_{B}{}^{C}{}_{D}
\widehat{\Omega}^{B}{}_{A}\widehat{\Omega}^{D}{}_{C},\label{eq60a}\\
T=T_{\rm tr}+T_{\rm int}&=&\frac{m}{2}g_{ij}v^{i}v^{j}+\frac{1}{2}\mathcal{R}^{i}{}_{j}{}^{k}{}_{l}
\Omega^{j}{}_{i}\Omega^{l}{}_{k},\label{eq61a}
\end{eqnarray}
where $\mathcal{L}^{A}{}_{B}{}^{C}{}_{D}$, $\mathcal{R}^{i}{}_{j}{}^{k}{}_{l}$ are both constant and symmetric in bi-indices $\left({}^{A}{}_{B},{}^{C}{}_{D}\right)$, $\left({}^{i}{}_{j},{}^{k}{}_{l}\right)$. There are no metrics which would be simultaneously left and right affinely invariant and non-degenerate. The reason is that the affine group is not semisimple and contains the normal subgroup composed of translations.

There exist however some models of the internal kinetic energy invariant simultaneously under the left and right action of the affine group. They are given by the linear combination of second-order Casimir invariants:
\begin{equation}\label{eq62a}
T_{\rm int}=\frac{A}{2}{\rm Tr}\left(\Omega^{2}\right)+\frac{B}{2}\left({\rm Tr}\ \Omega\right)^{2}=\frac{A}{2}{\rm Tr}\left(\widehat{\Omega}^{2}\right)+\frac{B}{2}\left({\rm Tr}\ \widehat{\Omega}\right)^{2},
\end{equation}
where $A$, $B$ are constants.

The $B$-term is a merely secondary correction. The main structure is controlled by the $A$-term. It is not positively definite and this property of the kinetic energy might seem embarrassing. It turns out however that at least for the incompressible body the negative contribution to $T_{\rm int}$ may encode the dynamics, without introducing any potential energy into Lagrangian or Hamiltonian. Such geodetic highly-invariant models may be often solved explicitly, analytically, in terms of some well-known special functions on groups. In some situations they may be solved in terms of exponential functions on groups, at least to some extent.

There are also other interesting models where the total kinetic energy, including translational one, is invariant under the left-acting isometry group and the right-acting total affine one. There are also models of opposite properties, i.e., invariant under the left-acting affine group and right-acting isometry one. The model affinely invariant on the right may be interpreted as a very drastic discretization of the Arnold description of ideal fluids \cite{Arn_78}. In Arnold theory this was the Hamiltonian system on the infinite-dimensional group of all volume-preserving diffeomorphisms; in our model this is the finite-dimensional group of affine volume-preserving mappings. Such models may be useful when describing molecules and droplets of nuclear matter. It is interesting that models affinely invariant in the physical space may be also realistic and convenient in condensed matter theory. The point is that due to strong interactions and strong concentration of matter, molecules may be non-sensitive (in their kinetic energy terms) to the metric $g$ of the physical space; instead, they may "feel", e.g., the Cauchy deformation tensor $C$ as a metric object underlying the structure of the kinetic energy. Incidentally, such situations are faced with, e.g., in solid state physics. The kinetic energy of electron is then based on the so-called tensor of effective mass, not on the usual metric geometry. Moreover, the tensor of effective mass may have various exotic properties, e.g., it need not be positively definite.

If we insist on positive definiteness, then it is interesting that when some phenomenological constants are appropriately chosen, then the left- or right-invariant affine kinetic energies of the total (internal together with translational) motion may have the positive signature. And at the same time they have the structure admitting rigorous analytical solutions.

The kinetic energy invariant under spatial isometries but not under the larger spatial group and simultaneously invariant under all material affine transformations has the following form: 
\begin{equation}\label{eq63a}
T=\frac{m}{2}g_{ij}v^{i}v^{j}+
\frac{I}{2}g_{ij}\Omega^{i}{}_{k}\Omega^{j}{}_{l}g^{kl}+
\frac{A}{2}{\rm Tr}\left(\Omega^{2}\right)+\frac{B}{2}\left({\rm Tr}\ \Omega\right)^{2}.
\end{equation}
The kinetic energy invariant under spatial affine group and under material isometries (but not under a larger group of material, i.e., Lagrangian transformations) has the shape as follows:
\begin{equation}\label{eq64}
T=\frac{m}{2}\eta_{AB}\widehat{v}^{A}\widehat{v}^{B}+
\frac{I}{2}\eta_{AB}\widehat{\Omega}^{A}{}_{C}\widehat{\Omega}^{B}{}_{D}\eta^{CD}+
\frac{A}{2}{\rm Tr}\left(\widehat{\Omega}^{2}\right)+\frac{B}{2}\left({\rm Tr}\ \widehat{\Omega}\right)^{2},
\end{equation}
where in both formulae above $m$, $I$, $A$, and $B$ are inertial constants.

Obviously, the two last terms in both formulae (\ref{eq63a}), (\ref{eq64}) are respectively equal to each other and different symbols $\Omega$, $\widehat{\Omega}$ are used only for aesthetic reasons. 

Let us remind the formula (\ref{eq53b}) and notice that (\ref{eq63a}), (\ref{eq64}) may be written in the following suggestive forms:
\begin{eqnarray}
T&=&\frac{m}{2}G_{AB}\widehat{v}^{A}\widehat{v}^{B}+
\frac{I}{2}G_{AB}\widehat{\Omega}^{A}{}_{C}\widehat{\Omega}^{B}{}_{D}G^{-1CD}+
\frac{A}{2}{\rm Tr}\left(\widehat{\Omega}^{2}\right)+
\frac{B}{2}\left({\rm Tr}\ \widehat{\Omega}\right)^{2},
\nonumber\label{eq65}\\
T&=&\frac{m}{2}C_{ij}v^{i}v^{j}+
\frac{I}{2}C_{ij}\Omega^{i}{}_{k}\Omega^{j}{}_{l}C^{-1kl}+
\frac{A}{2}{\rm Tr}\left(\Omega^{2}\right)+
\frac{B}{2}\left({\rm Tr}\ \Omega\right)^{2}.
\nonumber\label{eq66}
\end{eqnarray}
If we use Lagrangians of traditional potential forms, i.e.,
\begin{equation}\label{eq67}
L=T-V(x,\varphi),
\end{equation}
then the Legendre transformation may be written in any of the following equivalent forms, cf. also formulas (\ref{eq76+1})--(\ref{eq76+7}):
\begin{eqnarray}
p({\rm tr})_{i}=\frac{\partial L}{\partial v^{i}}=\frac{\partial T}{\partial v^{i}},&\quad& \Sigma^{j}{}_{i}=\frac{\partial L}{\partial \Omega^{i}{}_{j}}=\frac{\partial T}{\partial \Omega^{i}{}_{j}},\label{eq68}\\
\widehat{p}({\rm tr})_{A}=\frac{\partial L}{\partial \widehat{v}^{A}}=\frac{\partial T}{\partial \widehat{v}^{A}},&\quad& \widehat{\Sigma}^{B}{}_{A}=\frac{\partial L}{\partial \widehat{\Omega}^{A}{}_{B}}=\frac{\partial T}{\partial \widehat{\Omega}^{A}{}_{B}}.\label{eq69}
\end{eqnarray}
There are also various mixed possibilities.

Inverting these formulae, i.e., expressing velocities in terms of canonical momenta and substituting them to $T$, we obtain the kinetic terms $\mathcal{T}$ of Hamiltonians. We can admit also some potentials $V(x,\varphi)$ and consider the total Hamiltonians. However, it turns out that if the body is incompressible, the kinetic term alone may encode the dynamics of elastic vibrations. This resembles the Maupertuis principle where the dynamics (including some hidden version of the potential energy) may be encoded in the metric structure of the configuration space, i.e., in some geodetic model \cite{Arn_78}.

The mentioned model of incompressible body predicts both the bounded motion (below some energy threshold) and unbounded one. The only obstacle comes from compressibility, which must be stabilised by some potential term preventing the body from the dissociation (splitting) or contraction (collapse).

So, it is convenient to separate the isochoric motion from the pure compressibility. In the case of affinely-invariant dynamical models this is achieved by introducing on the real line $\mathbb{R}$ the centre of mass of the logarithmic deformation invariants,
\begin{equation}\label{eq70}
q=\frac{1}{n}\left(q^{1}+\cdots+q^{n}\right),
\end{equation}
and the corresponding conjugate momentum,
\begin{equation}\label{eq71}
p=p_{1}+\cdots+p_{n}.
\end{equation}
These quantities have to do with the uniform dilatations. This is rather exceptional phenomenon and should be prevented by some potential $V(q)$. As mentioned, other rotational and deformative modes of motion may be described in a satisfactory way by some geodetic dynamical models. Nevertheless, some potentials are also admissible. The most realistic of them are superpositions of binary and dilatational models:
\begin{equation}\label{eq72}
V\left(q^{1},\ldots,q^{n}\right)=V(q)+\frac{1}{2}\sum_{i\neq j}f_{ij}\left(
q^{i}-q^{j}\right),
\end{equation}
where the second term corresponds to the shear-like part of the motion.

The next quantities to be used are Casimir invariants built of $\Sigma$, $\widehat{\Sigma}$, i.e., 
\begin{equation}\label{eq73}
C(k)={\rm Tr}\left(\Sigma^{k}\right)={\rm Tr}\left(\widehat{\Sigma}^{k}\right),
\end{equation}
and of their trace-less parts, i.e.,
\begin{equation}\label{eq74}
C_{{\rm SL}(n)}(k)={\rm Tr}\left(\sigma^{k}\right)={\rm Tr}\left(\widehat{\sigma}^{k}\right),
\end{equation}
where
\begin{equation}\label{eq75}
\sigma=\Sigma-\frac{1}{n}\left({\rm Tr}\ \Sigma\right)\mathbb{I}_{n},\qquad 
\widehat{\sigma}=\widehat{\Sigma}-\frac{1}{n}\left({\rm Tr}\ \widehat{\Sigma}\right)\mathbb{I}_{n}.
\end{equation}
We need also the squared magnitudes of spin and vorticity,
\begin{equation}\label{eq76}
\|S\|^{2}=-\frac{1}{2}{\rm Tr}\left(S^{2}\right),\qquad \|V\|^{2}=-\frac{1}{2}{\rm Tr}\left(V^{2}\right),
\end{equation}
but {\bf ATTENTION:} those two positively-definite quantities, i.e., $\|S\|^{2}$ and $\|V\|^{2}$, are not equal to each other when some deformation occurs!

The model of $\mathcal{T}$ affinely invariant in space but metrically invariant in the system of particle (Lagrangian variables) has the following form:
\begin{equation}\label{eq77}
\mathcal{T}^{\rm aff-metr}_{\rm int}=\frac{1}{2\alpha}{\rm Tr}\left(\widehat{\Sigma}^{2}\right)+
\frac{1}{2\beta}\left({\rm Tr}\ \widehat{\Sigma}\right)^{2}+
\frac{1}{2\mu}\|V\|^{2},
\end{equation}
where
\begin{equation}\label{eq79}
\alpha=I+A,\quad \beta=-\frac{\left(I+A\right)\left(I+A+nB\right)}{B},\quad \mu=\frac{I^{2}-A^{2}}{I}.
\end{equation}
For the model of $T$ metrically invariant only in space but affinely invariant within the body we obtain that
\begin{equation}\label{eq78}
\mathcal{T}^{\rm metr-aff}_{\rm int}=\frac{1}{2\alpha}{\rm Tr}\left(\Sigma^{2}\right)+
\frac{1}{2\beta}\left({\rm Tr}\ \Sigma\right)^{2}+
\frac{1}{2\mu}\|S\|^{2}.
\end{equation}
In other words we have respectively
\begin{eqnarray}
\mathcal{T}^{\rm aff-metr}_{\rm int}&=&\frac{1}{2\alpha}C(2)+\frac{1}{2\beta}C(1)^{2}+
\frac{1}{2\mu}\|V\|^{2},\label{eq80}\\
\mathcal{T}^{\rm metr-aff}_{\rm int}&=&\frac{1}{2\alpha}C(2)+\frac{1}{2\beta}C(1)^{2}+
\frac{1}{2\mu}\|S\|^{2}.\label{eq81}
\end{eqnarray}
Separating the dilatational motion and shear, we obtain respectively
\begin{eqnarray}
\mathcal{T}^{\rm aff-metr}_{\rm int}&=&\frac{1}{2\left(I+A\right)}C_{{\rm SL}(n)}(2)+\frac{1}{2n\left(I+A+nB\right)}p^{2}\nonumber\\
&+&
\frac{I}{2\left(I^{2}-A^{2}\right)}\|V\|^{2},\label{eq82}\\
\mathcal{T}^{\rm metr-aff}_{\rm int}&=&\frac{1}{2\left(I+A\right)}C_{{\rm SL}(n)}(2)+\frac{1}{2n\left(I+A+nB\right)}p^{2}\nonumber\\
&+&
\frac{I}{2\left(I^{2}-A^{2}\right)}\|S\|^{2}.\label{eq83}
\end{eqnarray}
These expressions become the doubly affine invariant model when we put $I=0$, i.e.,
\begin{equation}\label{eq85}
\mathcal{T}^{\rm aff}_{\rm int}=\frac{1}{2A}C(2)-
\frac{B}{2A\left(A+nB\right)}C(1)^{2}.
\end{equation}
And similarly, they may be obtained from (\ref{eq85}) when we substitute $A\mapsto I+A$ and introduce in addition the mentioned $\|V\|^{2}$-, $\|S\|^{2}$-terms. 

Obviously, the geodetic system with Lagrangian given by the internal kinetic 
term (\ref{eq85}) is explicitly solvable in terms of matrix exponents. The corresponding geodetics are given by
\begin{equation}\label{eq120+1}
\varphi(t)=\exp\left(\Omega t\right)\varphi_{0}=\varphi_{0}\exp\left(\widehat{\Omega}t\right),
\end{equation}
where $\varphi_{0}$, $\Omega$, $\widehat{\Omega}$ are constants and 
\begin{equation}\label{eq120+2}
\widehat{\Omega}=\varphi^{-1}\Omega\varphi.
\end{equation}

As mentioned above, it is assumed here that the translational motion is neglected or dynamically independent on the internal/relative one. It is clear that $\varphi_{0}$ is the initial configuration at $t=0$ and $\Omega$, $\widehat{\Omega}$ are constant (thus, automatically initial) values of affine velocities (\ref{eq2}); this is the reason for notation used in (\ref{eq120+2}). Those initial conditions are completely arbitrary. The above general solution (\ref{eq120+1}) is a priori obvious. However, we are interested mainly in dynamically relevant quantities $q^{i}$, $p_{i}$, $M^{a}{}_{b}$, $N^{a}{}_{b}$. Their time dependence may be in principle extracted from the exponential formula (\ref{eq120+1}), although it is not an easy task; some implicit function and inverse function procedures are meant. But on the level of differential equations for the relevant quantities $q^{i}$, $p_{i}$, $M^{a}{}_{b}$, $N^{a}{}_{b}$ this would be hopeless. There is also some qualitative advantage from using the exponential description of geodetics. Namely, on its basis one can show that there are open subsets of bounded and unbounded trajectories ("below threshold" and "above threshold" motions) in the configuration space. And of course, it is so also on the level of the subfamily of state variables $(q^{i},M^{a}{}_{b},N^{a}{}_{b})$, because motion in $L,R$-variables is always bounded; the submanifolds of corresponding degrees of freedom are compact. And the details of motion in $L,R$-variables are not of large physical interest in problems like the scattering data, radiation and absorption. There are obvious quantum counterparts of these statements.

For Lagrangians (\ref{eq82}), (\ref{eq83}) the general solution is not any longer given by the exponential formula (\ref{eq120+1}). Only some special "stationary solutions" of such a form (obvious generalizations of stationary rotations in rigid body dynamics) do exist. In those solutions $\varphi_{0}$ is arbitrary, but  $\Omega$, $\widehat{\Omega}$ must satisfy certain algebraic conditions. So, for (\ref{eq82}), (\ref{eq83}) respectively the following must hold:
\begin{equation}\label{eq120+3}
\left[\widehat{\Omega},\widehat{\Omega}^{\eta T}\right]=0,\qquad \left[\Omega,\Omega^{g T}\right]=0,
\end{equation}
where the $\eta$- and $g-$transposes are respectively given by
\begin{equation}\label{eq120+4}
\left(\widehat{\Omega}^{\eta T}\right)^{A}{}_{B}=\eta^{AC}\eta_{BD}
\widehat{\Omega}^{D}{}_{C},\qquad \left(\Omega^{g T}\right)^{i}{}_{j}=g^{ik}g_{jl}\Omega^{l}{}_{k}.
\end{equation}
Roughly speaking, the corresponding affine velocities are "normal" in the $\eta$-sense for (\ref{eq82}) and in the $g$-sense for (\ref{eq83}).

Nevertheless, on the level of quantities $(q^{i},M^{a}{}_{b},N^{a}{}_{b})$
the exponential formula is still useful and the time dependence of those variables may be, as for (\ref{eq85}), extracted from it. Again one must remember that in the exponential procedure based on (\ref{eq85}), $A$ must be replaced by $(I+A)$. And everything concerning the bounded and non-bounded trajectories remains valid.

Let us now quote the one-dimensional lattice aspects of the classical dynamics \cite{D.48,D.49}. They are interesting in themselves and are strongly related to the hyperbolic version of Sutherland integrable lattices (at least in the geodetic case). In the affine-affine version they have the following form:
\begin{equation}\label{eq87}
\mathcal{T}^{\rm aff}_{\rm int}=\frac{1}{2\alpha}\sum_{a}p^{2}_{a}+
\frac{1}{32\alpha}\sum_{a\neq b}\frac{\left(M^{a}{}_{b}\right)^{2}}{{\rm sh}^{2}\frac{q^{a}-q^{b}}{2}}-\frac{1}{32\alpha}\sum_{a\neq b}\frac{\left(N^{a}{}_{b}\right)^{2}}{{\rm ch}^{2}\frac{q^{a}-q^{b}}{2}}+
\frac{1}{2\beta}p^{2}.
\end{equation}
The reasoning which leads from (\ref{eq62a}) to (\ref{eq87}) is a direct replication of that presented in formulas (\ref{eq76+1})--(\ref{eq76+7}), but now with the kinetic energy (\ref{eq62a}) used instead of (\ref{eq60}).

In the explicit binary representation:
\begin{eqnarray}
\mathcal{T}^{\rm aff}_{\rm int}&=&\frac{1}{4n\alpha}\sum_{a\neq b}\left(p_{a}-p_{b}\right)^{2}+
\frac{1}{32\alpha}\sum_{a\neq b}\frac{\left(M^{a}{}_{b}\right)^{2}}{{\rm sh}^{2}\frac{q^{a}-q^{b}}{2}}\nonumber\\
&-&\frac{1}{32\alpha}\sum_{a\neq b}\frac{\left(N^{a}{}_{b}\right)^{2}}{{\rm ch}^{2}\frac{q^{a}-q^{b}}{2}}+
\frac{n\alpha+\beta}{2n\alpha\beta}p^{2},\label{eq88}
\end{eqnarray}
or in a more apparent way:
\begin{eqnarray}
\mathcal{T}^{\rm aff}_{\rm int}&=&\frac{1}{4nA}\sum_{a\neq b}\left(p_{a}-p_{b}\right)^{2}+
\frac{1}{32A}\sum_{a\neq b}\frac{\left(M^{a}{}_{b}\right)^{2}}{{\rm sh}^{2}\frac{q^{a}-q^{b}}{2}}
\nonumber\\
&-&\frac{1}{32A}\sum_{a\neq b}\frac{\left(N^{a}{}_{b}\right)^{2}}{{\rm ch}^{2}\frac{q^{a}-q^{b}}{2}}+
\frac{1}{2n\left(A+nB\right)}p^{2}.\label{eq89}
\end{eqnarray}
The lattice structure of the dynamics of deformation invariants is obvious.

For the affine-metric and metric-affine kinetic energies we have respectively
\begin{eqnarray}
\mathcal{T}^{\rm aff-metr}_{\rm int}&=&\mathcal{T}^{\rm aff}_{\rm int}
\left[A\rightarrow I+A\right]+
\frac{I}{2\left(I^{2}-A^{2}\right)}\|V\|^{2},\label{eq90a}\\
\mathcal{T}^{\rm metr-aff}_{\rm int}&=&\mathcal{T}^{\rm aff}_{\rm int}
\left[A\rightarrow I+A\right]+
\frac{I}{2\left(I^{2}-A^{2}\right)}\|S\|^{2},\label{eq90b}
\end{eqnarray}
where $\mathcal{T}^{\rm aff}_{\rm int}\left[A\rightarrow I+A\right]$ denotes (\ref{eq85}) with $A$ replaced by $I+A$.

It is interesting that the Casimir invariant $C(2)$ has the following form for the incompressible body:
\begin{eqnarray}
C_{{\rm SL}(n)}(2)={\rm Tr}\left(\sigma^{2}\right)={\rm Tr}\left(\widehat{\sigma}^{2}\right)&=&\frac{1}{2n}\sum_{a\neq b}\left(p_{a}-p_{b}\right)^{2}\label{eq91}\\
&+&
\frac{1}{16}\sum_{a\neq b}\frac{\left(M^{a}{}_{b}\right)^{2}}{{\rm sh}^{2}\frac{q^{a}-q^{b}}{2}}
-\frac{1}{16}\sum_{a\neq b}\frac{\left(N^{a}{}_{b}\right)^{2}}{{\rm ch}^{2}\frac{q^{a}-q^{b}}{2}},\nonumber
\end{eqnarray}
neither $q$ nor $p$ enters here (no dilatational contribution to dynamics).

The mentioned possibility of encoding the dynamics of bounded elastic vibrations in affinely-invariant kinetic energy forms (without using any potential energy) is explicitly visualized in formulas (\ref{eq87}), (\ref{eq88}), (\ref{eq89}). Namely, the negative configuration describes the apparently exotic "centrifugal attraction" of deformation invariants. The total formulas describes the balance of attraction and repulsion which results in the existence of open families of bounded and escaping trajectories (open as subsets of the general solution, i.e., family of all trajectories).

It is seen that the deformation invariants $q^{a}$ behave like indistinguishable "particles" on the real line $\mathbb{R}$. On the quantum level they become some strange parastatistical "particles". This follows from the very peculiar non-uniqueness of the two-polar decomposition, as described above. Deformation invariants, as seen in the above formulas, interact mutually and are coupled to each other via the quantities $M^{a}{}_{b}$, $N^{a}{}_{b}$ which play the role of some "springs", respectively repulsive and attractive ones. These coupling "constants" are however variable; they are dynamical quantities which together with invariants $q^{a}$ satisfy the closed system of evolution equations obtained from (\ref{eq52}) by substituting our $\mathcal{T}_{\rm int}$ for the Hamiltonian $H$ and our $q^{a}$, $p_{a}$, $M^{a}{}_{b}$, $N^{a}{}_{b}$ for the state variable $F$. Incidentally, the situation becomes particularly suggestive in the planar case $n=2$, i.e., in the "Flatland" world as described by Abbott \cite{D.58}. This has to do with the exceptional properties of ${\rm GL}(2,\mathbb{R})$ among all groups ${\rm GL}(n,\mathbb{R})$. These properties follow from the fact that ${\rm SO}(2,\mathbb{R})$, being a one-dimensional Lie group, is Abelian. Because of this, $\widehat{\chi}=\chi$, $\widehat{\vartheta}=\vartheta$, $\widehat{\varrho}=\varrho=S$, $\widehat{\tau}=\tau=V$ and any of these matrices has only one independent component. Obviously, the same concerns the matrices $M$, $N$ and we shall use the following symbols:
\begin{equation}\label{eq91+1}
m:=M^{1}{}_{2},\qquad n:=N^{1}{}_{2}.
\end{equation}
The two-polar decomposition takes on the following matrix form:
\begin{equation}\label{eq91+2}
\varphi=LDR^{-1}=\left[
\begin{array}{cc}
\cos\alpha & -\sin\alpha \\
\sin\alpha & \cos\alpha
\end{array}\right]
\left[
\begin{array}{cc}
Q^{1} & 0 \\
0 & Q^{2}
\end{array}\right]\left[
\begin{array}{cc}
\cos\beta & \sin\beta \\
-\sin\beta & \cos\beta
\end{array}\right]
\end{equation}
and as usual, in affine models the exponential representation
\begin{equation}\label{eq91+3}
Q^{a}=\exp\left(q^{a}\right),\qquad a=1,2
\end{equation}
is convenient. Canonical momenta conjugate to $Q^{a}$ or $q^{a}$ are also denoted by the usual symbols $P_{a}$, $p_{a}$.

Canonical momenta conjugate to angular variables $\alpha$, $\beta$ are denoted by $p_{\alpha}$, $p_{\beta}$. One can show that, obviously,
\begin{eqnarray}\label{eq91+4}
\chi^{1}{}_{2}=\widehat{\chi}^{1}{}_{2}=-\frac{d\alpha}{dt},&\quad& \vartheta^{1}{}_{2}=\widehat{\vartheta}^{1}{}_{2}=-\frac{d\beta}{dt},\\
\varrho^{1}{}_{2}=\widehat{\varrho}^{1}{}_{2}=p_{\alpha},&\quad& \tau^{1}{}_{2}=\widehat{\tau}^{1}{}_{2}=p_{\beta}.
\end{eqnarray}
It is clear that
\begin{equation}\label{eq91+5}
m=p_{\beta}-p_{\alpha},\qquad n=p_{\beta}+p_{\alpha}.
\end{equation}
It is also convenient to introduce the variables $q$, $x$, characterizing respectively dilatation and shear, 
\begin{equation}\label{eq91+6}
q=\frac{1}{2}\left(q^{1}+q^{2}\right),\qquad x=q^{2}-q^{1}.
\end{equation}
The corresponding conjugate momenta $p$, $p_{x}$ are given by
\begin{equation}\label{eq91+7}
p=p_{1}+p_{2},\qquad p_{x}=\frac{1}{2}\left(p_{2}-p_{1}\right).
\end{equation}
Then for (\ref{eq87}) we obtain that
\begin{equation}\label{eq91+8}
\mathcal{T}^{\rm aff}_{\rm int}=\frac{p^{2}_{x}}{A}+
\frac{\left(p_{\beta}-p_{\alpha}\right)^{2}}{16A{\rm sh}^{2}\frac{x}{2}}-
\frac{\left(p_{\beta}+p_{\alpha}\right)^{2}}{16A{\rm ch}^{2}\frac{x}{2}}+
\frac{p^{2}}{4\left(A+2B\right)}
\end{equation}
and similarly (\ref{eq82}), (\ref{eq83}) become respectively
\begin{eqnarray}
\mathcal{T}^{\rm aff-metr}_{\rm int}&=&\mathcal{T}^{\rm aff}_{\rm int}
\left[A\rightarrow I+A\right]+
\frac{Ip^{2}_{\beta}}{I^{2}-A^{2}},\label{eq91+9}\\
\mathcal{T}^{\rm metr-aff}_{\rm int}&=&\mathcal{T}^{\rm aff}_{\rm int}
\left[A\rightarrow I+A\right]+
\frac{Ip^{2}_{\alpha}}{I^{2}-A^{2}},\label{eq91+10}
\end{eqnarray}
where, as previously, $\mathcal{T}^{\rm aff}_{\rm int}
\left[A\rightarrow I+A\right]$ denotes $\mathcal{T}^{\rm aff}_{\rm int}$ with $A$ replaced by $I+A$.

It is seen that the $x$-terms of the last three formulas may be written as
\begin{equation}\label{eq91+11}
\mathcal{T}^{\rm aff}_{\rm int}[x]=\frac{p^{2}_{x}}{A}+V^{\rm eff}_{m,n}(x),
\end{equation}
where the effective potential $V^{\rm eff}_{m,n}$ is given by
\begin{equation}\label{eq91+12}
V^{\rm eff}_{m,n}(x)=\frac{m^{2}}{16A{\rm sh}^{2}\frac{x}{2}}-
\frac{n^{2}}{16A{\rm ch}^{2}\frac{x}{2}}.
\end{equation}

Obviously, $\alpha$, $\beta$ are cyclic variables, thus, $p_{\alpha}$, $p_{\beta}$ are constants of motion and their fixed values $m$, $n$ of $\left(p_{\beta}-p_{\alpha}\right)$, $\left(p_{\beta}+p_{\alpha}\right)$ characterize the families of the $x$-motions in the general solutions. It is seen that if $|m|<|n|$, then at large $|x|$-distances the attractive negative term prevails; if $|m|>|n|$, then the repulsive term prevails for $|x|\rightarrow\infty$. In the first case one deals with bounded vibrations in the $x$-variable, i.e., in the mutual displacements of deformation invariants. As the first, repulsive term of $V^{\rm eff}_{m,n}(x)$ is positively-infinite when $x\rightarrow 0$, we are dealing formally with the characteristic shape known from the theory of intermolecular forces (cf., e.g., Fig. 1 and Fig. 2 with $x$ taken instead of $R$). If $|m|>|n|$, then $V^{\rm eff}_{m,n}(x)$ is purely repulsive. Thus, in the isochoric, ${\rm SL}(2,\mathbb{R})$-ruled part of motion we are dealing with some thresholds $|m|=|n|$ separating the bounded (oscillatory) and non-bounded (decaying) motions. What concerns the $q$-part, i.e., pure dilatations, motion is unbounded, just the uniform motion with constant velocity, thus, collapsing or infinitely expanding on the level of the variable $\exp(q)$. This is just the earlier mentioned necessity of correcting $\mathcal{T}^{\rm aff}_{\rm int}$ by some extra potential $V(q)$ stabilizing dilatations (the potential well or something very quickly increasing with $|q|$) \cite{D.28}. Here, in the exceptional case $n=2$ everything is easily seen because $M$, $N$ are now constants of motion. Everything remains true for $\mathcal{T}^{\rm aff-metr}_{\rm int}$ and $\mathcal{T}^{\rm metr-aff}_{\rm int}$, because $p_{\alpha}$, $p_{\beta}$ are constants of motion. As well everything remains valid on the quantized level, where one deals respectively with the discrete and continuous spectrum of energy.

All is so nice in affinely-invariant models. In traditional "d'Alembert" model (\ref{eq63}) only pure repulsion occurs in geodetic models and they are completely non-viable without introducing of some potential term. In particular, in two dimensions $n=2$, (\ref{eq63}) becomes
\begin{equation}\label{eq91+13}
\mathcal{T}_{\rm int}=\frac{1}{2I}\left(P^{2}_{1}+P^{2}_{2}\right)++V^{\rm eff}_{m,n}\left(Q^{1},Q^{2}\right),
\end{equation}
where
\begin{equation}\label{eq91+14}
V^{\rm eff}_{m,n}\left(Q^{1},Q^{2}\right)=
\frac{m^{2}}{4I\left(Q^{1}-Q^{2}\right)^{2}}-
\frac{n^{2}}{4I\left(Q^{1}+Q^{2}\right)^{2}}.
\end{equation}
Without the extra potential term, only the unbounded purely scattering motion of deformation invariants is possible.

It is very interesting to deviate for a while from the hyperbolic Sutherland lattices (\ref{eq91}) to their trigonometric analogues. Namely, if we consider a dynamical system on the unitary group ${\rm U}(n)$ and use again the analogue of the two-polar decomposition (\ref{eq31}),
\begin{equation}\label{eq92}
\varphi=LDR^{-1},
\end{equation}
where $L,R\in{\rm SO}(n,\mathbb{R})$ are real-orthogonal and $D$ is diagonal with complex unitary entries on the diagonal,
\begin{equation}\label{eq93}
D_{aa}=\exp\left(iq^{a}\right),\qquad q^{a}\in\mathbb{R},
\end{equation}
then for the combination of Casimir invariants controlled by constants $A$, $B$ we obtain the kinetic energy
\begin{equation}\label{eq94}
T_{\rm int}=-\frac{A}{2}{\rm Tr}\left(\Omega^{2}\right)-
\frac{B}{2}\left({\rm Tr}\ \Omega\right)^{2}=-\frac{A}{2}{\rm Tr}\left(\widehat{\Omega}^{2}\right)-
\frac{B}{2}\left({\rm Tr}\ \widehat{\Omega}\right)^{2},
\end{equation}
where $A>0$, $B>0$ and the Lie-algebraic elements
\begin{equation}\label{eq95}
\Omega=\frac{d\varphi}{dt}\varphi^{-1},\qquad \widehat{\Omega}=\varphi^{-1}\frac{d\varphi}{dt}=\varphi^{-1}\Omega\varphi
\end{equation}
are skew-Hermitian.

The corresponding kinetic energy becomes then in the Hamiltonian representation as follows:
\begin{eqnarray}
\mathcal{T}_{\rm int}&=&\frac{1}{2A}\sum_{a}p^{2}_{a}+
\frac{1}{32A}\sum_{a\neq b}\frac{\left(M^{a}{}_{b}\right)^{2}}{\sin^{2}\frac{q^{a}-q^{b}}{2}}\nonumber\\
&+&
\frac{1}{32A}\sum_{a\neq b}\frac{\left(N^{a}{}_{b}\right)^{2}}{\cos^{2}\frac{q^{a}-q^{b}}{2}}-
\frac{B}{2A\left(A+nB\right)}p^{2}.\label{eq96}
\end{eqnarray}
Separating explicitly the dilatational $(q,p)$-variables we obtain
\begin{eqnarray}
\mathcal{T}_{\rm int}&=&\frac{1}{4nA}\sum_{a\neq b}\left(p_{a}-p_{b}\right)^{2}+
\frac{1}{32A}\sum_{a\neq b}\frac{\left(M^{a}{}_{b}\right)^{2}}{\sin^{2}\frac{q^{a}-q^{b}}{2}}
\nonumber\\
&+&\frac{1}{32A}\sum_{a\neq b}\frac{\left(N^{a}{}_{b}\right)^{2}}{\cos^{2}\frac{q^{a}-q^{b}}{2}}+
\frac{1}{2n\left(A+nB\right)}p^{2}.\label{eq97}
\end{eqnarray}
The Casimir invariant $C(2)$ for ${\rm SU}(n)$ takes on the following form:
\begin{equation}\label{eq98}
C_{{\rm SU}(n)}(2)=\frac{1}{2n}\sum_{a\neq b}\left(p_{a}-p_{b}\right)^{2}+
\frac{1}{16}\sum_{a\neq b}\frac{\left(M^{a}{}_{b}\right)^{2}}{\sin^{2}\frac{q^{a}-q^{b}}{2}}
+\frac{1}{16}\sum_{a\neq b}\frac{\left(N^{a}{}_{b}\right)^{2}}{\cos^{2}\frac{q^{a}-q^{b}}{2}}.
\end{equation}

Let us notice an important role of the negative contribution to (\ref{eq91}). It describes attraction and stabilises the vibrating regime of incompressible bodies. All contributions to (\ref{eq98}) are positive. However, it does not matter in view of the circular topology of the $q$-variable and all $q^{a}$-ones in general. Because then it is impossible to distinguish between repulsion and attraction.

It is very interesting that in (\ref{eq90a}), (\ref{eq90b}) the characteristic $\|S\|^{2}$- and $\|V\|^{2}$-terms appear. Those invariants of spatial and material rotations are very suggestive and resemble certain formulae appearing in the description of Raman scattering, rotational and deformative excitation of nuclear matter, especially on the quantized level.

Let us also mention, one can suspect some physical interpretation in collective models, where both the $\|S\|^{2}$- and $\|V\|^{2}$-terms appear. They would contain highly symmetric affinely-invariant expressions controlled by $A$, $B$ as above and in addition two orthogonal terms restricting the symmetry group to isometries in the physical and material space (in Euler and Lagrange variables respectively). In terms of velocities they would be given by the following phenomenological formulae:
\begin{equation}\label{eq99}
T_{\rm int}=\frac{I_{1}}{2}g_{ik}g^{jl}\Omega^{i}{}_{j}\Omega^{k}{}_{l}+
\frac{I_{2}}{2}\eta_{KL}\eta^{MN}\widehat{\Omega}^{K}{}_{M}
\widehat{\Omega}^{L}{}_{N}+
\frac{A}{2}{\rm Tr}\left(\widehat{\Omega}^{2}\right)+
\frac{B}{2}\left({\rm Tr}\ \widehat{\Omega}\right)^{2}.
\end{equation}
Obviously, the third and fourth terms may be as well written as
\begin{equation}\label{eq100}
\frac{A}{2}{\rm Tr}\left(\Omega^{2}\right)+
\frac{B}{2}\left({\rm Tr}\ \Omega\right)^{2}.
\end{equation}
The first term may be alternatively written in the following form:
\begin{equation}\label{eq101}
\frac{I_{1}}{2}G_{KL}G^{-1MN}\widehat{\Omega}^{K}{}_{M}
\widehat{\Omega}^{L}{}_{N}.
\end{equation}
Similarly, the second one may be written as
\begin{equation}\label{eq102}
\frac{I_{2}}{2}C_{ik}C^{-1jl}\Omega^{i}{}_{j}\Omega^{k}{}_{l}.
\end{equation}
Every of these forms may appeal to some intuitions and may be suggestive when properly read out.

It is clear that after the Legendre transformation based on (\ref{eq99}) we obtain the following kinetic term of the Hamiltonian:
\begin{eqnarray}
\mathcal{T}_{\rm int}&=&
\frac{1}{2\overline{I_{1}}}g_{ik}g^{jl}\Sigma^{i}{}_{j}\Sigma^{k}{}_{l}+
\frac{1}{2\overline{I_{2}}}\eta_{KL}\eta^{MN}\widehat{\Sigma}^{K}{}_{M}
\widehat{\Sigma}^{L}{}_{N}\nonumber\\
&+&
\frac{1}{2\overline{A}}{\rm Tr}\left(\widehat{\Sigma}^{2}\right)+
\frac{1}{2\overline{B}}\left({\rm Tr}\ \widehat{\Sigma}\right)^{2},\label{eq103}
\end{eqnarray}
where $\overline{I_{1}}$, $\overline{I_{2}}$, $\overline{A}$, $\overline{B}$ are some inertial constants built algebraically of $I_{1}$, $I_{2}$, $A$, $B$. Let us also mention expressions similar to (\ref{eq100}), (\ref{eq101}), (\ref{eq102}).

It is also clear that the corresponding expressions for (\ref{eq103}) will contain on equal footing both terms proportional to $\|S\|^{2}$ and $\|V\|^{2}$. Let us also mention that the dynamics for $q^{a}$, $p_{a}$, $M^{a}{}_{b}$, $N^{a}{}_{b}$ is closed (at least when the potentials $V$ do not exist or depend only on the deformations invariants $q^{1},\ldots,q^{n}$, first of all on $q$ separated from $q^{a}-q^{b}$). Equations of motion written in terms of quantities $q^{a}$, $p_{a}$, $M^{a}{}_{b}$, $N^{a}{}_{b}$ are autonomous and independent of $L,R$-variables. They have the following form based on the Poisson brackets:
\begin{eqnarray}
\frac{dq^{a}}{dt}&=&\{q^{a},H\}=\frac{\partial H}{\partial p_{a}},\label{eq104aa}\\ \frac{dp_{a}}{dt}&=&\{p_{a},H\}=-\frac{\partial H}{\partial q^{a}},
\label{eq104ab}\\
\frac{dM^{a}{}_{b}}{dt}&=&\{M^{a}{}_{b},H\}=
\{M^{a}{}_{b},M^{c}{}_{d}\}\frac{\partial H}{\partial M^{c}{}_{d}}+\{M^{a}{}_{b},N^{c}{}_{d}\}\frac{\partial H}{\partial N^{c}{}_{d}},\label{eq104b}\\
\frac{dN^{a}{}_{b}}{dt}&=&\{N^{a}{}_{b},H\}=
\{N^{a}{}_{b},M^{c}{}_{d}\}\frac{\partial H}{\partial M^{c}{}_{d}}+\{N^{a}{}_{b},N^{c}{}_{d}\}\frac{\partial H}{\partial N^{c}{}_{d}},\label{eq104c}
\end{eqnarray}
where
\begin{equation}\label{eq105}
\{q^{a},p_{b}\}=\delta^{a}{}_{b},\quad \{q^{a},M^{c}{}_{d}\}=\{p_{a},M^{c}{}_{d}\}=\{q^{a},N^{c}{}_{d}\}=
\{p_{a},N^{c}{}_{d}\}=0.
\end{equation}
To obtain the Poisson brackets for $M$, $N$ one must use the brackets for $\widehat{\varrho}$, $\widehat{\tau}$ from which $M$, $N$ are built. In turn, the brackets for $\widehat{\varrho}$, $\widehat{\tau}$ are implied by the commutation relations in the Lie algebra ${\rm SO}(n,\mathbb{R})^{\prime}$ for ${\rm SO}(n,\mathbb{R})$ because $\widehat{\varrho}$, $\widehat{\tau}$ are Hamiltonian generators of transformations acting on $L$, $R$ in (\ref{eq32}), respectively, the following ones:
\begin{equation}\label{eq106}
\left[L^{i}{}_{a}\right]\mapsto\left[L^{i}{}_{b}A^{b}{}_{a}\right],\qquad 
\left[R^{K}{}_{a}\right]\mapsto\left[R^{K}{}_{b}B^{b}{}_{a}\right],
\end{equation}
where $A,B\in{\rm SO}(n,\mathbb{R})$. Those Hamiltonian generators are related to spin and vorticity variables as follows:
\begin{equation}\label{eq107}
\widehat{\varrho}^{a}{}_{b}=L^{-1a}{}_{i}S^{i}{}_{j}L^{j}{}_{b},\qquad \widehat{\tau}^{a}{}_{b}=-R^{-1a}{}_{K}V^{K}{}_{L}R^{L}{}_{b}.
\end{equation}
Therefore, after the Kronecker-delta shift of indices we obtain that
\begin{eqnarray}
\{\widehat{\varrho}_{ab},\widehat{\varrho}_{cd}\}&=&
\widehat{\varrho}_{ad}\delta_{cb}-\widehat{\varrho}_{cb}\delta_{ad}+
\widehat{\varrho}_{db}\delta_{ac}-\widehat{\varrho}_{ac}\delta_{db},
\label{eq108a}\\
\{\widehat{\tau}_{ab},\widehat{\tau}_{cd}\}&=&
\widehat{\tau}_{ad}\delta_{cb}-\widehat{\tau}_{cb}\delta_{ad}+
\widehat{\tau}_{db}\delta_{ac}-\widehat{\tau}_{ac}\delta_{db},
\label{eq108b}\\
\{\widehat{\varrho}_{ab},\widehat{\tau}_{cd}\}&=&0,
\label{eq108c}
\end{eqnarray}
and finally,
\begin{eqnarray}
\{M_{ab},M_{cd}\}=\{N_{ab},N_{cd}\}&=&
M_{cb}\delta_{ad}-M_{ad}\delta_{bc}+M_{ac}\delta_{db}-M_{db}\delta_{ac},\qquad
\label{eq109a}\\
\{M_{ab},N_{cd}\}&=&
N_{cb}\delta_{ad}-N_{ad}\delta_{bc}+N_{ac}\delta_{db}-N_{db}\delta_{ac}.
\label{eq109b}
\end{eqnarray}
Solving (in principle) equations of motion (\ref{eq104aa}), (\ref{eq104ab}), (\ref{eq104b}), (\ref{eq104c}) and performing partially the inverse Legendre transformation, one obtains that
\begin{equation}\label{eq110}
\widehat{\chi}^{a}{}_{b}=\frac{\partial H}{\partial \widehat{\varrho}^{b}{}_{a}},\qquad \widehat{\vartheta}^{a}{}_{b}=\frac{\partial H}{\partial \widehat{\tau}^{b}{}_{a}},
\end{equation}
so, one can find (in principle) the time dependence of $\widehat{\chi}$, $\widehat{\vartheta}$ and then the time dependence of $L$, $R$ may be found by solving the following equation:
\begin{equation}\label{eq111}
\frac{dL}{dt}=L\widehat{\chi},\qquad \frac{dR}{dt}=R\widehat{\vartheta}.
\end{equation}

\section{Quantum models}

We are interested in studying physical phenomena in the micro- and nano-scale, where the quantized theory must be used.

Roughly speaking, for quantum models the configuration space of internal/relative degrees of freedom may be identified with ${\rm L}^{2}\left({\rm GL}(n,\mathbb{R})\right)$, a bit more rigorously with ${\rm L}^{2}\left({\rm LI}(U,V)\right)$, where ${\rm LI}(U,V)$ denotes the manifold of linear isomorphisms of $U$ onto $V$ (usually the volume-preserving ones). Kinetic energies are usually based on some underlying metrics,
\begin{equation}\label{eq112}
\Gamma_{\mu\nu}(y)dy^{\mu}\otimes dy^{\nu},\qquad T=\frac{1}{2}\Gamma_{\mu\nu}(y)\frac{dy^{\mu}}{dt}\frac{dy^{\nu}}{dt}.
\end{equation}
The corresponding Hilbert space of wave functions ${\rm L}^{2}\left(Q,\mu\right)$ consists of square-integrable functions on the configuration space $Q$ endowed with the canonical measure $\mu$ built of the metric tensor $\Gamma$. The corresponding scalar product is given by
\begin{equation}\label{eq113}
\langle\Psi_{1},\Psi_{2}\rangle=\int\overline{\Psi}_{1}(y)\Psi_{2}(y)d\mu(y),
\end{equation}
where
\begin{equation}\label{eq114-115}
d\mu(y)=\sqrt{|\det\left[\Gamma_{\mu\nu}(y)\right]|}dy^{1}\cdots dy^{N},\qquad N=\dim Q.
\end{equation}
Fortunately, our metrics have certain invariance properties and because of this the measures $\mu$ are what mathematicians call Haar measures \cite{Loo_53}. Those are invariant under the left (Euler-spatial) and right (Lagrangian-material) translations, cf. (\ref{eq18}). If we take into account the translational degrees of freedom, then the corresponding measure $\alpha$ is given by
\begin{eqnarray}
d\alpha(x,\varphi)&=&\left(\det\varphi\right)^{-n-1}dx^{1}\cdots dx^{n}d\varphi^{1}{}_{1}\cdots d\varphi^{n}{}_{n}\nonumber\\
&=&
\left(\det\varphi\right)^{-1}d\lambda(\varphi)dx^{1}\cdots dx^{n},\label{eq116}\\
d\lambda(\varphi)&=&\left(\det\varphi\right)^{-n}d\varphi^{1}{}_{1}\cdots d\varphi^{n}{}_{n}.
\label{eq117}
\end{eqnarray}
The two-polar (singular value) decomposition implies that
\begin{equation}\label{eq118}
d\lambda(\varphi)=d\lambda(L,q,R)=\prod_{i\neq j}\left|{\rm sh}\left(q^{i}-q^{j}\right)\right|d\mu(L)d\mu(R)dq^{1}\cdots dq^{n},
\end{equation}
where $\mu$ denotes the Haar measures on the compact, connected and simply connected orthogonal groups ${\rm SO}(n,\mathbb{R})$. 

If we wish to consider the incompressible bodies, then the dilatation factor must be cancelled by the corresponding Dirac distribution,
\begin{equation}\label{eq119}
d\lambda_{{\rm SL}(n)}(\varphi)=\prod_{i\neq j}\left|{\rm sh}\left(q^{i}-q^{j}\right)\right|d\mu(L)d\mu(R)
\delta\left(q^{1}+\cdots+q^{n}\right)dq^{1}\cdots dq^{n}.
\end{equation}
When quantizing the "d'Alembert-like" models (\ref{eq63}), it is more convenient to deal with the usual Lebesgue measure $\ell$ on the manifold of internal degrees of freedom,
\begin{equation}\label{eq120}
d\ell(\varphi)=d\varphi^{1}{}_{1}\cdots d\varphi^{n}{}_{n}.
\end{equation}
Then, if we use the two-polar splitting, we have that
\begin{equation}\label{eq121}
d\ell(L,Q,R)=P_{\ell}\left(Q^{1},\ldots,Q^{n}\right)
d\mu(L)d\mu(R)dQ^{1}\cdots dQ^{n},
\end{equation}
where 
\begin{equation}\label{eq122}
P_{\ell}\left(Q^{1},\ldots,Q^{n}\right)=\prod_{a\neq b}\left|\left(Q^{a}\right)^{2}-\left(Q^{b}\right)^{2}\right|=\prod_{a\neq b}\left|\left(Q^{a}+Q^{b}\right)\left(Q^{a}-Q^{b}\right)\right|.
\end{equation}

If translational degrees of freedom are explicitly taken into account, then in analogy to (\ref{eq116}) we have that
\begin{equation}\label{eq123}
da(x,\varphi)=d\ell(\varphi)dx^{1} \cdots \ dx^{n}=d\ell\left(L,Q,R\right)dx^{1} \cdots dx^{n}.
\end{equation}
It will be also convenient to write the Haar measure (\ref{eq118}) in the abbreviated form analogous to (\ref{eq121}), (\ref{eq122}):
\begin{equation}\label{eq124}
d\lambda(\varphi)=d\lambda(L,q,R)=P_{\lambda}\left(q^{1},\ldots,q^{n}\right)
d\mu(L)d\mu(R)dq^{1}\cdots dq^{n},
\end{equation}
where now
\begin{equation}\label{eq125}
P_{\lambda}\left(q^{1},\ldots,q^{n}\right)=\prod_{i\neq j}\left|{\rm sh}\left(q^{i}-q^{j}\right)\right|.
\end{equation}

Obviously, the measure $\lambda$ is invariant under all transformations (\ref{eq18}); similarly, $\alpha$ is invariant under (\ref{eq18}) accompanied by all affine mappings acting on $x^{i}$, i.e., on translational degrees of freedom. Unlike this, the Lebesgue measures $\ell$, $a$ are invariant only if the mentioned mappings are restricted to isometries.

In general, the procedure of Schr\"odinger quantization in a Riemannian manifold $\left(Q,\Gamma\right)$ begins from introducing the operator of kinetic energy, proportional to the Laplace-Beltrami operator \cite{Dav_58,D.27},
\begin{equation}\label{eq126}
{\bf T}=-\frac{\hbar^{2}}{2}\Delta_{\Gamma}=-\frac{\hbar ^{2}}{2}\Gamma^{\mu\nu}\nabla_{\mu}\nabla_{\nu}=\frac{1}{2}
\Gamma^{\mu\nu}\left(\frac{\hbar}{i}\nabla_{\mu} \right)\left(\frac{\hbar}{i}\nabla_{\nu}\right),
\end{equation}
where $\nabla_{\mu}$ denotes the operator of the covariant differentiation in the $\Gamma$-Levi-Civita sense along the $\mu$-th coordinate axis. 

The operators $(\hbar/i)\nabla_{\mu}$ and $-(\hbar^{2}/2)\Delta$ are formally self-adjoint, i.e., satisfy \cite{Dav_58,D.27}
\begin{equation}\label{eq127}
\left\langle{\bf A}\Psi_{1}|\Psi_{2}\right\rangle=\left\langle\Psi_{1}|{\bf A} \Psi_{2}\right\rangle,
\end{equation}
if $\Psi_{1}$, $\Psi_{2}$ are confined to some dense subdomain of ${\rm L}^{2}(Q,\mu)$ consisting of sufficiently smooth functions. Being differential operators, they are evidently unbounded. They are formally self-adjoint, because the Levi-Civita parallel transport does preserve the Riemann measure $\mu$.

However, in general such a procedure would be extremely strenuous. It would be very difficult to avoid mistakes and even if avoiding them we would obtain some rather obscure, non-readable expressions. Fortunately, differential operators generating left and right regular translations in the configuration space enable one to simplify the procedure in a remarkable way.

Namely, it may be easily shown that transformations of wave functions induced by the argumentwise action of (\ref{eq19}) are unitary in ${\rm L}^{2}\left({\rm LI}(U,V),\lambda\right)$, just due to the left and right invariance of the measure $\lambda$. Similarly, the usual vector translations in the physical space, just as (\ref{eq19}) themselves, are unitary in ${\rm L}^{2}(Q,\alpha)$. On the other hand, there is a unitary failure in  ${\rm L}^{2}\left({\rm LI}(U,V),\ell\right)$, ${\rm L}^{2}(Q,a)$, unless the argumentwise action of transformations on $\Psi$ will be accompanied by certain multiplicative correction factor built of the determinants of $A$, $B$ in (\ref{eq19}).

Because of all that, the operators
\begin{eqnarray}
{\bf \Sigma}^{a}{}_{b}&:=&\frac{\hbar}{i}{\bf L}^{a}{}_{b}=\frac{\hbar}{i}\varphi^{a}{}_{K}\frac{\partial}{\partial \varphi^{b}{}_{K}},\label{eq128a}\\
\widehat{\bf \Sigma}^{A}{}_{B}&:=&\frac{\hbar}{i}{\bf R}^{A}{}_{B}=\frac{\hbar}{i}\varphi^{a}{}_{B}\frac{\partial}{\partial \varphi^{a}{}_{A}}=\varphi^{-1A}{}_{a}\varphi^{b}{}_{B}{\bf \Sigma}^{a}{}_{b}
\label{eq128b}
\end{eqnarray}
are "formally Hermitian". The "formally anti-Hermitian" first-order differential operators ${\bf L}^{a}{}_{b}$, ${\bf R}^{A}{}_{B}$ are infinitesimal generators of the mentioned unitary groups in  ${\rm L}^{2}(Q,\alpha)$. Obviously, the operators of spin and vorticity, i.e.,
\begin{eqnarray}
{\bf S}^{a}{}_{b}&=&{\bf \Sigma}^{a}{}_{b}-{\bf \Sigma}_{b}{}^{a}={\bf \Sigma}^{a}{}_{b}-g^{ac}g_{bd}{\bf \Sigma}^{d}{}_{c},\label{eq129a}\\
{\bf V}^{A}{}_{B}&=&{\widehat {\bf \Sigma}}^{A}{}_{B}-{\widehat {\bf \Sigma}}_{B}{}^{A}={\widehat {\bf \Sigma}}^{A}{}_{B}-\eta^{AC}\eta_{BD}{\widehat {\bf \Sigma}}^{D}{}_{C},\label{eq129b}
\end{eqnarray}
or rather their $\left(i/\hbar\right)$-multiplies, are infinitesimal generators of "spatial" and "material" rotations. ${\bf \Sigma}^{a}{}_{b}$, ${\widehat{\bf \Sigma}}^{A}{}_{B}$ are operators representing affine spin with respect to the space- and body-fixed axes respectively. An important point is that no problems of ordering operators appear here. Namely, just due to the geometric interpretation of operators as generators of well-defined transformation groups, the ordering is exactly like in (\ref{eq128a}), (\ref{eq128b}).

Instead of fighting with formulae like (\ref{eq126}) we simply write the following well-defined expression for the model of internal kinetic energy affinely invariant both in the Euler and Lagrange variables:
\begin{eqnarray}\label{eq130}
{\bf T}^{\rm aff-aff}_{\rm int}&=&\frac{1}{2A}{\bf \Sigma}^{i}{}_{j}{\bf \Sigma}^{j}{}_{i}-\frac{B}{2A(A+nB)}{\bf \Sigma}^{i}{}_{i}{\bf \Sigma}^{j}{}_{j}\nonumber\\
&=&\frac{1}{2A}{\widehat{\bf \Sigma}}^{K}{}_{L}{\widehat {\bf \Sigma}}^{L}{}_{K}
-\frac{B}{2A(A+nB)}{\widehat {\bf \Sigma}}^{K}{}_{K}{\widehat {\bf \Sigma}}^{L}{}_{L},
\end{eqnarray}
just the automatic replacement of $\Sigma^{i}{}_{j}$, ${\widehat \Sigma}^{K}{}_{L}$ by the operators ${\bf \Sigma}^{i}{}_{j}$, ${\widehat {\bf \Sigma}}^{K}{}_{L}$ in the corresponding classical formulae.

For the models only isometrically invariant in Euler variables and affinely invariant in Lagrange ones we obtain obviously the following operator of kinetic energy:
\begin{equation}\label{eq131}
{\bf T}^{\rm metr-aff}_{\rm int}=\frac{1}{2\widetilde{I}}g_{ik}g^{jl}{\bf \Sigma}^{i}{}_{j}{\bf \Sigma}^{k}{}_{l}
+\frac{1}{2\widetilde{A}}{\bf \Sigma}^{i}{}_{j}{\bf \Sigma}^{j}{}_{i}+\frac{1}{2\widetilde{B}}
{\bf \Sigma}^{i}{}_{i}{\bf \Sigma}^{j}{}_{j},
\end{equation}
where
\begin{equation}\label{eq132}
\widetilde{I}=\frac{I^{2}-A^{2}}{I}, \qquad \widetilde{A}=\frac{A^{2}-I^{2}}{A}, \qquad \widetilde{B}=-\frac{\left(I+A\right)\left(I+A+nB\right)}{B}. 
\end{equation}

Obviously, the kinetic energy operator affinely invariant in Euler variables and only isometrically invariant in Lagrange ones can be obtained in the following dual form:
\begin{equation}\label{eq133}
{\bf T}^{\rm aff-metr}_{\rm int}=\frac{1}{2\widetilde{I}}\eta_{AB}\eta^{CD}{\widehat {\bf \Sigma}}^{A}{}_{C}
{\widehat {\bf \Sigma}}^{B}{}_{D}
+\frac{1}{2\widetilde{A}}{\widehat {\bf \Sigma}}^{A}{}_{B}{\widehat {\bf \Sigma}}^{B}{}_{A}+\frac{1}{2\widetilde{B}}
{\widehat {\bf \Sigma}}^{A}{}_{A}{\widehat {\bf \Sigma}}^{B}{}_{B},
\end{equation}
with the same meaning of inertial constants (\ref{eq132}).

The accompanying expressions for the operators of translational kinetic energy are given respectively by
\begin{eqnarray}
{\bf T}^{\rm metr-aff}_{\rm tr}&=&\frac{1}{2m}g^{ij}{\bf p}({\rm tr})_{i}{\bf p}({\rm tr})_{j}=
\frac{1}{2m}G^{-1AB}{\bf {\widehat p}}({\rm tr})_{A}{\bf {\widehat p}}({\rm tr})_{B},\label{eq134}\\
{\bf T}^{\rm aff-metr}_{\rm tr}&=&\frac{1}{2m}C^{-1ij}{\bf p}({\rm tr})_{i}{\bf p}({\rm tr})_{j}=
\frac{1}{2m}\eta^{AB}{\bf {\widehat p}}({\rm tr})_{A}{\bf {\widehat p}}({\rm tr})_{B},\label{eq135}
\end{eqnarray}
where, let us remind, ${\bf p}({\rm tr})_{i}$, ${\bf {\widehat p}}({\rm tr})_{A}$ are formally Hermitian operators of linear (translational) momentum as expressed in spatial (laboratory)/material (co-moving) terms:
\begin{equation}\label{eq136}
{\bf  p}({\rm tr})_{a}= \frac{\hbar}{i }\frac{\partial}{\partial x^{a}}, \qquad \widehat{\bf  p}({\rm tr})_{K}=\varphi^{a}{}_{K}{\bf p}_{a}= \frac{\hbar}{i }\varphi^{a}{}_{K}\frac{\partial}{\partial x^{a}}.
\end{equation}

On the classical level we used the logarithmic dilatational invariant $q$ (\ref{eq70}) and its conjugate canonical momentum $p$ (\ref{eq71}). In quantized theory this momentum is represented by the following formally Hermitian operator:
\begin{equation}\label{eq139}
{\bf p}= \frac{\hbar}{i}\frac{\partial}{\partial q}.
\end{equation}
It is interrelated to the shear parts of the affine spin (deviator) through the following formulae:
\begin{equation}\label{eq140}
{\bf s}^{a}{}_{b}= {\bf \Sigma}^{a}{}_{b}-\frac{1}{n}{\bf p}\delta^{a}{}_{b}, \qquad \widehat{\bf s}^{A}{}_{B}= {\bf {\widehat \Sigma}}^{A}{}_{B}-\frac{1}{n}{\bf p}\delta^{A}{}_{B}, \qquad {\bf p}={\bf \Sigma}^{a}{}_{a}={\bf {\widehat \Sigma}}^{A}{}_{A}.
\end{equation}

Just like on the classical level, one can perform a partial separation of shear (incompressible motion) and dilatations. Expressions for the operators of internal kinetic energy become then as follows:
\begin{eqnarray}
{\bf T}^{\rm aff-aff}_{\rm int}&=&\frac{1}{2A}{\bf C}_{{\rm SL}(n)}(2)+\frac{1}{2n(A+nB)}{\bf p}^{2},\label{eq141}\\
{\bf T}^{\rm metr-aff}_{\rm int}&=&\frac{1}{2(I+A)}{\bf C}_{{\rm SL}(n)}(2)\nonumber\\
&+&\frac{1}{2n(I+A+nB)}{\bf p}^{2}+\frac{I}{2\left(I^{2}-A^{2}\right)}\|{\bf S}\|^{2},\label{eq142}\\
{\bf T}^{\rm aff-metr}_{\rm int}&=&\frac{1}{2(I+A)}{\bf C}_{{\rm SL}(n)}(2)\nonumber\\
&+&\frac{1}{2n(I+A+nB)}{\bf p}^{2}+\frac{I}{2\left(I^{2}-A^{2}\right)}\|{\bf V}\|^{2},\label{eq143}
\end{eqnarray}
where on the quantum level we mean that
\begin{equation}\label{eq144}
{\bf C}_{{\rm SL}(n)}(k)={\bf s}^{a}{}_{b}{\bf s}^{b}{}_{c} \cdots {\bf s}^{r}{}_{s}{\bf s}^{s}{}_{a}={\bf {\widehat s}}^{A}{}_{B}{\bf {\widehat s}}^{B}{}_{C} \cdots {\bf {\widehat s}}^{R}{}_{S}{\bf {\widehat s}}^{S}{}_{A},
\end{equation}
(k factors) and
\begin{equation}\label{eq145}
\|{\bf S}\|^{2}:=-\frac{1}{2}{\bf S}^{a}{}_{b}{\bf S}^{b}{}_{a}, \qquad \|{\bf V}\|^{2}:=-\frac{1}{2}{\bf V}^{A}{}_{B}{\bf V}^{B}{}_{A}.
\end{equation}
Sometimes, however, it is more convenient to write simply that
\begin{eqnarray}
{\bf T}^{\rm aff-aff}_{\rm int}&=&\frac{1}{2A}{\bf C}(2)-\frac{B}{2A(A+nB)}{\bf p}^{2},\label{eq146}\\
{\bf T}^{\rm metr-aff}_{\rm int}&=&\frac{1}{2\alpha}{\bf C}(2)+\frac{1}{2\beta}{\bf p}^{2}+\frac{1}{2\mu}\|{\bf S}\|^{2},\label{eq147}\\
{\bf T}^{\rm aff-metr}_{\rm int}&=&\frac{1}{2\alpha}{\bf C}(2)+\frac{1}{2\beta}{\bf p}^{2}+\frac{1}{2\mu}\|{\bf V}\|^{2},\label{eq148}
\end{eqnarray}
where $\alpha$, $\beta$, $\mu$ are given by (\ref{eq79}) and again the operators ${\bf C}(k)$ are built according to the Casimir prescription,
\begin{equation}\label{eq149}
{\bf C}(k)={\bf \Sigma}^{a}{}_{b}{\bf \Sigma}^{b}{}_{c} \cdots {\bf \Sigma}^{r}{}_{s}{\bf \Sigma}^{s}{}_{a}={\bf {\widehat\Sigma}}^{A}{}_{B}{\bf {\widehat\Sigma}}^{B}{}_{C} \cdots {\bf {\widehat\Sigma}}^{R}{}_{S}{\bf {\widehat\Sigma}}^{S}{}_{A},
\end{equation}
$k$ multiplicative factors.

The quantized version of the model (\ref{eq99}), (\ref{eq103}) will be based on the following operator of kinetic energy:
\begin{eqnarray}\label{eq150}
{\bf T}^{\rm metr-metr}_{\rm int}&=&\frac{1}{2\overline{I}_{1}}g_{ik}g^{jl}{\bf \Sigma}^{i}{}_{j}{\bf \Sigma}^{k}{}_{l}+\frac{1}{2 \overline{I}_{2}}\eta_{KL}\eta^{MN}{\bf {\widehat \Sigma}}^{K}{}_{M}{\bf {\widehat \Sigma}}^{L}{}_{N}\nonumber\\
&+&\frac{1}{2 \overline A}{\bf {\widehat \Sigma}}^{K}{}_{L}{\bf {\widehat \Sigma}}^{L}{}_{K}+\frac{1}{2 \overline B}{\bf {\widehat \Sigma}}^{K}{}_{K}{\bf {\widehat \Sigma}}^{L}{}_{L}.
\end{eqnarray}

In all the above expressions for ${\bf T}$ there is no problem with the ordering of operators just due to the geometric interpretation of the operators ${\bf \Sigma}^{i}{}_{j}$, ${\bf {\widehat \Sigma}}^{A}{}_{B}$ as generators of some groups of unitary transformations. And it is easily seen that all the resulting expressions for ${\bf T}$ automatically are formally Hermitian.
      
What  concerns (\ref{eq150})  itself, this model is spatially and materially invariant only under the isometry groups. However, it has a nice structure because it is a superposition of two affinely-invariant terms and two additional ones which restrict this symmetry to a weaker one, namely isometric. Manipulations with phenomenological inertial constants enable one to control somehow those symmetry properties. After some calculations performed on (\ref{eq150}), we obtain that
\begin{equation}\label{eq151}
{\bf T}^{\rm metr-metr}_{\rm int}=\frac{1}{2\overline {\alpha}}{\bf C}(2)+\frac{1}{2 \overline{\beta}}{\bf p}^{2}+\frac{1}{2\overline{\mu}}\|{\bf S}\|^{2}+\frac{1}{2\overline{\nu}}\|{\bf V}\|^{2},
\end{equation}
where $\overline{\alpha}$, $\overline{\beta}$, $\overline{\mu}$, $\overline{\nu}$ are some constants built of $\overline {I_{1}}$, $\overline {I_{2}}$, $\overline {A}$, $\overline {B}$, in analogy to (\ref{eq79}). It may be also convenient to represent the above expression as
\begin{equation}\label{eq152}
{\bf T}^{\rm metr-metr}_{\rm int}=\frac{1}{2a}{\bf C}_{{\rm SL}(n)}(2)+\frac{1}{2 b}{\bf p}^{2}+\frac{1}{2c}\|{\bf S}\|^{2}+\frac{1}{2d}\|{\bf V}\|^{2},
\end{equation}
where $a$, $b$, $c$, $d$ are some new constants. In any case, the formula (\ref{eq152}) may be also postulated as something primary, without the intermediary step (\ref{eq150}), just as a natural generalisation/unification of (\ref{eq147}), (\ref{eq148}).

Just as in the classical theory, spin and vorticity operators may be expressed in terms of their components with respect to bases co-moving with the $L$- and $R$-gyroscopes,
\begin{equation}\label{eq153}
{\bf \widehat{r}}^{a}{}_{b}=L^{-1a}{}_{i}L^{j}{}_{b}{\bf S}^{i}{}_{j}, \qquad {\bf \widehat{t}}^{a}{}_{b}=-R^{-1a}{}_{K}R^{L}{}_{b}{\bf V}^{K}{}_{L},
\end{equation}
the ordering of variables as indicated. Due to the orthogonality of $L$, $R$ it is clear that the following holds for the "magnitudes":
\begin{equation}\label{eq154}
{\bf \widehat{r}}^{a}{}_{b}{\bf \widehat{r}}^{b}{}_{a}={\bf S}^{i}{}_{j}{\bf S}^{j}{}_{i}, \quad {\bf \widehat{t}}^{a}{}_{b}{\bf \widehat{t}}^{b}{}_{a}={\bf V}^{K}{}_{L}{\bf V}^{L}{}_{K}.
\end{equation}
In geodetic cases, or with potentials $V\left(q^{1},\ldots,q^{n}\right)$ built of deformation invariants, ${\bf S}^{i}{}_{j}$, ${\bf V}^{A}{}_{B}$ are quantum constants of motion, i.e., they commute with the Hamiltonian ${\bf H}={\bf T}+V$. It is not the case with ${\bf r}$, ${\bf t}$, however their squared magnitudes, being equal to those of ${\bf S}$, ${\bf V}$, are also constants of motion.

Just like in the classical theory, in certain quantum expressions it is convenient to use the following operators:
\begin{equation}\label{eq155}
{\bf M}^{a}{}_{b}=-{\bf \widehat{r}}^{a}{}_{b}-{\bf \widehat{t}}^{a}{}_{b}, \qquad {\bf N}^{a}{}_{b}={\bf \widehat{r}}^{a}{}_{b}-{\bf \widehat{t}}^{a}{}_{b},
\end{equation}
which enable one to perform a "partial diagonalization" of the kinetic energy.

It is clear that for all geodetic models or more general dynamical models with potentials depending only on deformation invariants, the eigenvalues of orthogonal Casimirs of spin and vorticity
\begin{equation}\label{eq156}
\|{\bf S}\|^{2}=\|{\bf \widehat{r}}\|^{2},\qquad \|{\bf V}\|^{2}=\|{\bf \widehat{t}}\|^{2}
\end{equation}
are "good" quantum numbers. In the physical three-dimensional case they are given respectively by \cite{D.51,D.50,D.34}
\begin{equation}\label{eq157}
C\left({\bf S},s\right)=\hbar^{2}s(s+1),\qquad C\left({\bf V},j\right)=\hbar^{2}j(j+1),
\end{equation}
where $s$, $j$ are non-negative integers,
\begin{equation}\label{eq158}
s=0,1,2,\ldots,\qquad j=0,1,2,\ldots\qquad \left(s,j \in \left\{ 0 \right\} \cup \mathbb{N}\right).
\end{equation}
It may be also shown \cite{JJS_Denver,JJS-Kov_04,all_04,all_05} that $s$, $j$ may be non-negative integers and half-integers,
\begin{equation}\label{eq159}
s=0,\frac{1}{2},1,\frac{3}{2},\ldots,\qquad j=0,\frac{1}{2},1,\frac{3}{2},\ldots\qquad \left(s,j \in \left\{ 0 \right\} \cup \frac{1}{2}\mathbb{N}\right),
\end{equation}
with the condition, however, that $s$, $j$ are simultaneously integer or half-integer, i.e., $(j-s)$ is an integer:
\begin{equation}\label{eq160}
\left|j-s\right|=0,1,2,\ldots\qquad \left(\left|j-s\right| \in \left\{ 0 \right\} \cup \mathbb{N}\right).
\end{equation}
This has to do with admitting some special kind of multivalued wave functions; the procedure suggested among others by Pauli and Reiss \cite{Pauli,Reiss}. The configuration space of internal degrees of freedom, originally identified with  ${\rm
GL}^{+}(3,\mathbb{R})$, is then replaced by its universal covering manifold, i.e., the universal covering group $\overline{{\rm
GL}^{+}(3,\mathbb{R})}$, which, by the way, is not a linear group (it does not admit any realization in terms of finite-dimensional matrices).

There is an interesting message of formulae (\ref{eq142}), (\ref{eq143}), (\ref{eq147}), (\ref{eq148}), (\ref{eq151}), (\ref{eq152}), concerning the spectrum of radiation of objects described by ${\bf T}_{\rm int}$ as Hamiltonians of internal motion. More generally, this applies also to Hamiltonians of the following form:
\begin{equation}\label{eq161}
{\bf H}={\bf T}+V\left(q^{1},\ldots,q^{n}\right),
\end{equation}
i.e., ones with potential terms depending only on deformation invariants (scalar stretchings). Those spectral peculiarities appear, in particular, in phenomena like the Raman scattering, when the absorbed light gives rise to the internal excited states which decay through radiation proving the splitting of internal energy levels. This splitting is imposed onto some system of background levels corresponding to the spectra of the first two terms in (\ref{eq142}), (\ref{eq143}), (\ref{eq147}), (\ref{eq148}), (\ref{eq151}), (\ref{eq152}) and is described by the terms proportional to operators $\|{\bf S}\|^{2}$, $\|{\bf V}\|^{2}$. As we know, in the three-dimensional physical space these operators have spectra (\ref{eq157}), (\ref{eq158}), under certain conditions (\ref{eq159}),  (\ref{eq160}).

In expressions (\ref{eq142}), (\ref{eq147}) we easily recognise the rotational Raman splitting controlled by the quantity $\hbar^{2}s(s+1)$. These terms correspond exactly to excitation of rotations in the physical space.

From this point of view the models (\ref{eq143}), (\ref{eq148}) describe something else, although the splitting has again the structure $\hbar^{2}j(j+1)$. But this is not the quantized rotation. Instead, it is some part of the quantized deformative motion, i.e., some aspect of quantized deformations. The "rotational" expression $\hbar^{2}j(j+1)$ is simply due to the rotation of squeezing, rotation of the deformation tensor. So, in spite of the  $\hbar^{2}j(j+1)$-structure this is not any rotation of the "molecule" in space, this is rather something like the rotation of some external factors suppressing the "molecule".

In (\ref{eq151}), (\ref{eq152}) one deals simultaneously with both aspects: the quantized $\hbar^{2}s(s+1)$-rotation in space and the quantized $\hbar^{2}j(j+1)$-controlled deformation process. This might be something realistic, because in spectra of some micro-objects one observes splittings of the $\hbar^{2}j(j+1)$-type which cannot be interpreted as a quantized rotation in space.

In nuclear physics there appear terms of the type $\hbar^{2}I(I+1)$, where $I$ is an isospin. It is so even in elementary particles, where the mass formula of eight-fold way, obtained by Gell-Mann and Okubo for hadrons reads:
\begin{equation}\label{eq162}
M=a+bY+c\left( I(I+1)-\frac{1}{4}Y^{2}\right),
\end{equation}
where $I$ denotes the isospin quantum number, $Y$ is so-called hypercharge, and $a$, $b$, $c$ are constants. This is particularly remarkable when we consider the model (\ref{eq96}), (\ref{eq97}), where  ${\rm
GL}(n,\mathbb{R})$ is replaced by $U(n)$ is such a way that deformation invariants $\exp\left(q^{a}\right)$ are "compactified" to $\exp\left(iq^{a}\right)$.

It seems that the invariance structure and symmetry groups are so fundamental for dynamics that they may lead to quite similar models in rather mutually remote areas of physical phenomena.

Incidentally, let us mention that the term linear in $Y$ may have an analogue within our treatment, and namely if we admit in the formula for the kinetic energy some terms linear in generalized velocities. The only geometrically invariant ones are those proportional to ${\rm Tr}\ \Omega={\rm Tr}\ {\widehat \Omega}$, i.e., proportional to the operator ${\bf p}$ on the quantum level. Apparently, the terms proportional to velocities might seem exotic. Let us remind, however, that they appear in analytical mechanics of charged particles moving in the magnetic field. We did not consider above such models with kinetic terms linear in velocities, nevertheless, it may be easily done.

Let us quote for the sake of completeness some formulae concerning the quantum description. In $n$ dimensions our wave functions may be expanded in the series
\begin{equation}\label{eq163}
\Psi(\varphi)=\Psi(L,q,R)=\sum_{\alpha,\beta\in\Theta}\sum_{m,n=1}^{N(\alpha)}
\sum_{k,l=1}^{N(\beta)}\mathcal{D}^{\alpha}_{mn}(L)
f^{\alpha\beta}_{{}^{nk}_{ml}}(q)
\mathcal{D}^{\beta}_{kl}\left(R^{-1}\right)
\end{equation}
with the following meaning of symbols:
\begin{itemize}
\item $\Theta$ is the set of equivalence classes of unitary irreducible representations of ${\rm SO}(n,\mathbb{R})$,
\item $N(\alpha)$ is the dimension of the $\alpha$-th representation class; as ${\rm SO}(n,\mathbb{R})$ is compact, $N(\alpha)$ is finite.
\end{itemize}
This follows from the Peter-Weyl theorem \cite{D.51,D.50} applied to ${\rm SO}(n,\mathbb{R})$.

In the physical case $n=3$, $\Theta$ is in principle the set of non-negative integers, $\alpha$, $\beta$ are, just as above, denoted by $s,j=0,1,2,\ldots$, $N(s)=2s+1$, $N(j)=2j+1$, and the indices $(m,n)$, $(k,l)$ are considered as jumping by $1$ from $-s$ to $s$ and from $-j$ to $j$ respectively, and $\mathcal{D}^{s}$, $\mathcal{D}^{j}$ are standard expressions for unitary irreducible representations of ${\rm SO}(3,\mathbb{R})$. As mentioned, according to certain ideas by Pauli, it is possible to admit some two-valued wave functions $\Psi$, or more rigorously, wave functions defined on the covering group $\overline{{\rm GL}(n,\mathbb{R})}$. This group is nonlinear, i.e., non-realizable in terms of finite matrices.

As mentioned, $\alpha$, $\beta$ ($s$, $j$) are "good" quantum numbers, so it is often convenient to use just the reduced amplitudes
\begin{equation}\label{eq164}
\Psi^{\alpha\beta}_{ml}(\varphi)=\Psi^{\alpha\beta}_{ml}(L,q,R)=
\sum_{n=1}^{N(\alpha)}\sum_{k=1}^{N(\beta)}\mathcal{D}^{\alpha}_{mn}(L)
f^{\alpha\beta}_{nk}(q)\mathcal{D}^{\beta}_{kl}\left(R^{-1}\right).
\end{equation}

In the physical case $n=3$ this becomes
\begin{equation}\label{eq165}
\Psi^{sj}_{ml}(\varphi)=\Psi^{sj}_{ml}(L,q,R)=\sum_{n=-s}^{s}
\sum_{k=-j}^{j}\mathcal{D}^{s}_{mn}(L)f^{sj}_{nk}(q)
\mathcal{D}^{j}_{kl}\left(R^{-1}\right).
\end{equation}
As mentioned, when two-valued wave functions are admitted, then  ${\rm SO}(3,\mathbb{R})$ is to be replaced by  ${\rm SU}(2)$ and in the above series only such terms appear that $(j-s)$ is an integer. Obviously, the following eigenequations hold:
\begin{eqnarray}
\|{\bf S}\|^{2}\Psi^{sj}_{ml}&=&\|{\bf {\widehat{r}}}\|^{2}\Psi^{sj}_{ml}=\hbar^{2}s(s+1)\Psi^{sj}_{ml},
\label{eq166a}\\
\|{\bf V}\|^{2}\Psi^{sj}_{ml}&=&\|{\bf {\widehat{t}}}\|^{2}\Psi^{sj}_{ml}=\hbar^{2}j(j+1)\Psi^{sj}_{ml}.
\label{eq166b}
\end{eqnarray}
Traditionally one uses the convention that $m$, $l$ are related to the eigenvalues of ${\bf S}_{3}$, ${\bf V}_{3}$, the third components of spin and vorticity:
\begin{equation}\label{eq167}
{\bf S}_{3}\Psi^{sj}_{ml}=\hbar m\Psi^{sj}_{ml},\qquad 
{\bf V}_{3}\Psi^{sj}_{ml}=\hbar l\Psi^{sj}_{ml}.
\end{equation}
Similarly, when the values $n$, $k$ in the superposition (\ref{eq165}) are kept fixed and we retain only the corresponding single term, for the resulting reduced amplitudes we have that
\begin{equation}\label{eq168}
{\bf \widehat{r}}_{3}\Psi^{sj}_{{}^{ml}_{nk}}=\hbar n\Psi^{sj}_{{}^{ml}_{nk}}, \qquad {\bf {\widehat{t}}}_{3}\Psi^{sj}_{{}^{ml}_{nk}}=\hbar k\Psi^{sj}_{{}^{ml}_{nk}}.
\end{equation}

In three dimensions, when $\mathcal{D}^{s}_{mn}$ are well-known functions found explicitly by Wigner, this means that the dependence of $\Psi$ on "angular" variables $L$, $R$ is in a sense explicitly known. And the action of differential operators occurring in our formulae is "algebraized". In the two-dimensional space, when $n=2$, this is just the expansion into Fourier series. In three dimensions we have
\begin{equation}\label{eq169}
{\bf S}^{a}{}_{b}\Psi^{sj}=S^{sa}{}_{b}\Psi^{sj},\qquad 
{\bf V}^{A}{}_{B}\Psi^{sj}=\Psi^{sj}S^{jA}{}_{B},
\end{equation}
where $S^{s}$, $S^{j}$ are matrices of angular momenta indexed by $s$, $j$. In the academic general case $s$, $j$ would have to be replaced by some labels $\alpha$, $\beta$. If $n=3$, then $S^{s}$, $S^{j}$ are standard matrices $(2s+1)\times(2s+1)$, $(2j+1)\times(2j+1)$ which are explicitly known and quoted in any textbook on quantum mechanics, e.g., one by Landau and Lifshitz \cite{LanLif} (cf. also \cite{D.32,D.34}).

In explicitly matrix terms we can write that
\begin{equation}\label{eq170}
\Psi^{\alpha\beta}(L,q,R)=\mathcal{D}^{\alpha}(L)f^{\alpha\beta}(q)
\mathcal{D}^{\beta}\left(R^{-1}\right).
\end{equation}
Differential operators ${\bf \widehat{r}}$, ${\bf \widehat{t}}$ are algebraized as follows:
\begin{eqnarray}
{\bf \widehat{r}}^{a}{}_{b}\Psi^{\alpha\beta}&=&\mathcal{D}^{\alpha}(L)S^{\alpha a}{}_{b}f^{\alpha\beta}(q)\mathcal{D}^{\beta}\left(R^{-1}\right),\label{eq171}\\
{\bf \widehat{t}}^{a}{}_{b}\Psi^{\alpha\beta}&=&\mathcal{D}^{\alpha}(L)
f^{\alpha\beta}(q)S^{\beta a}{}_{b}\mathcal{D}^{\beta}\left(R^{-1}\right).\label{eq172}
\end{eqnarray}
In certain formulae it is convenient to use the symbols $\overrightarrow{S^{\alpha a}{}_{b}}$,
$\overleftarrow{S^{\beta a}{}_{b}}$, where 
\begin{equation}\label{eq173}
\overrightarrow{S^{\alpha a}{}_{b}}f^{\alpha\beta}:=S^{\alpha a}{}_{b}f^{\alpha\beta}, \qquad \overleftarrow{S^{\beta a}{}_{b}}f^{\alpha\beta}:=f^{\alpha\beta}S^{\beta a}{}_{b}.
\end{equation}
The Schr\"odinger equation reduces then to the infinite system of multicomponent Schr\"odinger equations for the reduced amplitudes given by the system of $N(\alpha)\times N(\beta)$, i.e., physically $(2s+1)\times (2j+1)$, $q$-dependent matrices $f^{\alpha\beta}(q)$. The scalar product $\langle\Psi_{1}|\Psi_{2}\rangle$ for the complete wave functions may be expressed in the following way through the one for reduced amplitudes:
\begin{eqnarray}
\langle\Psi_{1}|\Psi_{2}\rangle&=&\sum_{\alpha,\beta\in\Theta}\frac{1}{N(\alpha) N(\beta)}\int {\rm Tr}\left(f_{1}^{+\alpha\beta}\left(q^{1},\ldots,q^{n}\right)
f_{2}^{\alpha\beta}\left(q^{1},\ldots,q^{n}\right)\right)\nonumber\\
&\times&P_{\lambda}\left(q^{1},\ldots,q^{n}\right)dq^{1}\cdots dq^{n},\label{eq174}
\end{eqnarray}
where $P_{\lambda}\left(q^{1},\ldots,q^{n}\right)$ is given by (\ref{eq125}) and $N(\alpha)$, $N(\beta)$ are dimensions of irreducible representations $\alpha,\beta\in\Theta$. They are finite because ${\rm SO}(n,\mathbb{R})$ are compact groups.

Similarly, for the d'Alembert-type models one uses the representation:
\begin{eqnarray}
\langle\Psi_{1}|\Psi_{2}\rangle&=&\sum_{\alpha,\beta\in\Theta}\frac{1}{N(\alpha) N(\beta)}\int{\rm Tr}\left(f_{1}^{+\alpha\beta}\left(Q^{1},\ldots,Q^{n}\right)
f_{2}^{\alpha\beta}\left(Q^{1},\ldots,Q^{n}\right)\right)\nonumber\\
&\times&P_{\ell}\left(Q^{1},\ldots,Q^{n}\right)dQ^{1}\cdots dQ^{n},\label{eq175}
\end{eqnarray}
where $P_{\ell}\left(Q^{1},\ldots,Q^{n}\right)$ is given by (\ref{eq122}).

The operator (\ref{eq130}) for the doubly invariant kinetic energy may be written as follows:
\begin{eqnarray}
{\bf T}^{\rm aff-aff}_{\rm int}&=&
-\frac{\hbar^{2}}{2A}{\bf D}_{\lambda}+\frac{\hbar ^{2}B}{2 A\left(A+nB\right)}\frac{\partial^{2}}{\partial q^{2}}\nonumber\\
&+&\frac{1}{32A}\sum_{a,b}\frac{\left({\bf M}^{a}{}_{b}\right)^{2}}{{\rm sh}^{2}\frac{q^{a}-q^{b}}{2}}-\frac{1}{32A}\sum_{a,b}\frac{\left({\bf N}^{a}{}_{b}\right)^{2}}{{\rm ch}^{2}\frac{q^{a}-q^{b}}{2}},\label{eq176}
\end{eqnarray}
where the differential operator ${\bf D}_{\lambda}$ is given by
\begin{equation}\label{eq177}
{\bf D}_{\lambda}=\frac{1}{P_{\lambda}}\sum_{a}\frac{\partial}{\partial q^{a}}P_{\lambda}\frac{\partial}{\partial q^{a}}=\frac{1}{2a}\sum_{a}\frac{\partial^{2}}{\partial \left(q^{a}\right)^{2}}+\sum_{a}\frac{\partial\ln P_{\lambda}}{\partial q^{a}}\frac{\partial}{\partial q^{a}}.
\end{equation}

Let us observe that unlike to what might be naively expected, the operator ${\bf D}_{\lambda}$ involving differential operators $\partial/\partial q^{a}$ is not the usual $\mathbb{R}^{n}$ (physically $\mathbb{R}^{3}$) Laplace operator. One can reduce it to such a form by modifying the dependent variable,
\begin{equation}\label{eq178}
\Phi:=\sqrt{P_{\lambda}}\Psi,
\end{equation}
but the price is that an artificial amended potential appears:
\begin{equation}\label{eq179}
\widetilde{V}=-\frac{\hbar}{2A}\frac{1}{P_{\lambda}^{2}}+
\frac{\hbar^{2}}{4A}\frac{1}{P_{\lambda}}\sum_{a}
\left(\frac{\partial P_{\lambda}}{\partial q^{a}}\right)^{2}.
\end{equation}
As mentioned, the time-independent Schr\"odinger equation, i.e., eigenequation
\begin{equation}\label{eq180}
{\bf H}\Psi=E\Psi,
\end{equation}
splits into the infinite family of multicomponent amplitudes $f^{\alpha\beta}\left(q^{1},\ldots,q^{n}\right)$ involving, however, only $n$ independent variables $q^{1},\ldots,q^{n}$ instead the primary $n^{2}$ ones $\varphi^{i}{}_{A}$,
\begin{equation}\label{eq181}
{\bf H}^{\alpha\beta}f^{\alpha\beta}=E^{\alpha\beta}f^{\alpha\beta},
\end{equation}
where the reduced Hamiltonians ${\bf H}^{\alpha\beta}$ have the following form:
\begin{eqnarray}\label{eq182}
{\bf H}^{\alpha\beta}f^{\alpha\beta}&=&-\frac{\hbar^{2}}{2A}{\bf D}_{\lambda}f^{\alpha\beta}\nonumber\\
&+&\frac{1}{32A}\sum_{a,b}\frac{\left(\overleftarrow{S^{\beta a}{}_{b}}-\overrightarrow{S^{\alpha a}{}_{b}}\right)^{2}}{{\rm sh}^{2}\frac{q^{a}-q^{b}}{2}}f^{\alpha\beta}
-\frac{1}{32A}\sum_{a,b}\frac{\left(\overleftarrow{S^{\beta a}{}_{b}}+\overrightarrow{S^{\alpha a}{}_{b}}\right)^{2}}{{\rm ch}^{2}\frac{q^{a}-q^{b}}{2}}f^{\alpha\beta}\nonumber\\
&+&\frac{\hbar^{2}B}{2A\left(A+nB\right)}\frac{\partial^{2}f^{\alpha\beta} }{\partial q^{2}}+V\left(q^{1},\ldots,q^{n}\right)f^{\alpha\beta},
\end{eqnarray}
where the meaning of symbols $\overrightarrow{S^{\alpha a}{}_{b}}$, $\overleftarrow{S^{\beta a}{}_{b}}$ is like in (\ref{eq173}). The potential energy is assumed to depend merely on logarithmic deformation invariants $q^{1},\ldots,q^{n}$. As mentioned, on the level of incompressible motion, i.e., for ${\rm SL}(n,\mathbb{R})$, it is possible to remain on the purely geodetic level. Then it is sufficient to admit potentials $V(q)$ depending only on the dilatational parameter $q$ (\ref{eq70}). The problem splits then into the geodetic one on ${\rm SL}(n, \mathbb{R})$ and one-dimensional quantized oscillations in the $q$-variable. As usually one deals with almost incompressible (almost isochoric) motion, as for $V(q)$ some simple phenomenological model may be used, e.g., some potential well or "steep" oscillator. In metric-affine and affine-metric models, (\ref{eq182}) is respectively modified by the following terms:
\begin{equation}\label{eq184}
-\frac{\hbar^{2}}{2\mu}C\left(\alpha,2\right),\qquad -\frac{\hbar^{2}}{2\mu}C\left(\beta,2\right),
\end{equation}
based on second-order Casimir invariants for the orthogonal group ${\rm SO}(n,\mathbb{R})$. In the physical case $n=3$, they become \cite{LanLif}
\begin{equation}\label{eq185}
\frac{\hbar^{2}}{2\mu}s(s+1),\qquad \frac{\hbar^{2}}{2\mu}j(j+1)
\end{equation}
with the previous meaning of symbols $s$, $j$.

In more hypothetical metric-metric models (\ref{eq151}), (\ref{eq152}), we are dealing with the following additive correction term:
\begin{equation}\label{eq186}
\frac{\hbar^{2}}{2c}s(s+1)+\frac{\hbar^{2}}{2\alpha}j(j+1).
\end{equation}

Finally, let us quote a rather not very useful, although seemingly "more familiar", d'Alembert model:
\begin{eqnarray}
{\bf H}^{\alpha\beta}f^{\alpha\beta}&=&-\frac{\hbar^{2}}{2I}{\bf D}_{\ell}f^{\alpha\beta}+\frac{1}{8I}\sum_{a,b}
\frac{\left(\overleftarrow{S^{\beta a}{}_{b}}-\overrightarrow{S^{\alpha a}{}_{b}}\right)^{2}}{\left(Q^{a}-Q^{b}\right)^{2}}f^{\alpha\beta}\nonumber\\
&+&\frac{1}{8I}\sum_{a,b}
\frac{\left(\overleftarrow{S^{\beta a}{}_{b}}+\overrightarrow{S^{\alpha a}{}_{b}}\right)^{2}}{\left(Q^{a}+Q^{b}\right)^{2}}f^{\alpha\beta}+
V\left(Q^{1},\ldots,Q^{n}\right)f^{\alpha\beta},
\label{eq187}
\end{eqnarray}
where
\begin{equation}\label{eq188}
{\bf D}_{\ell}=\frac{1}{P_{\ell}}\sum_{a}\frac{\partial}{\partial Q^{a}}P_{\ell}\frac{\partial}{\partial Q^{a}}=
\sum_{a}\frac{\partial^{2}}{\partial \left(Q^{a}\right)^{2}}+
\frac{1}{P_{\ell}}\sum_{a}\left(
\frac{\partial P_{\ell}}{\partial Q^{a}}\right)^{2}.
\end{equation}
Let us note that for (\ref{eq187}) both the classical and quantum geodetic models would be completely non-physical and the use of some relatively general potential $V\left(Q^{1},\ldots,Q^{n}\right)$ is absolutely unavoidable.

It is worth to mention that in planar problems, when $n=2$, there exists a wide class of models which are both qualitatively physical and integrable \cite{Gol_03,Gol_04a,Gol_04b,Mart_04a,Mart_04b,Mart_08,all_04,all_05}.

\section{Some qualitative remarks}

It is well known that for typical complex objects like molecules, the structure of Raman spectra depends strongly on the mutual positions and splittings of excited energy levels of internal motion. In molecules a typical picture is as follows, cf. some pictures below based on \cite{Marg_69,Slat_68,Slat_74}:
\begin{itemize}
\item[$(i)$] The main background is created by the system of electronic energy levels. Usually they are analysed and approximately calculated on the basis of the Born-Oppenheimer approximation \cite{Marg_69,Slat_68,Slat_74}. In principle the separation of those levels is such that the corresponding quantum transitions result in radiation of visible light.
\item[$(ii)$] Those levels are, as a matter of fact, bands consisting of systems of vibrational (deformative) energy levels. The frequencies of radiative transitions within those bands are placed within the visible light and near infrared ranges.
\item[$(iii)$] And finally, the vibrational levels split into the rotational ones. Here the resulting frequencies belong to the far infrared and radio ranges.
\end{itemize}
\begin{center}
\includegraphics[scale=0.25]{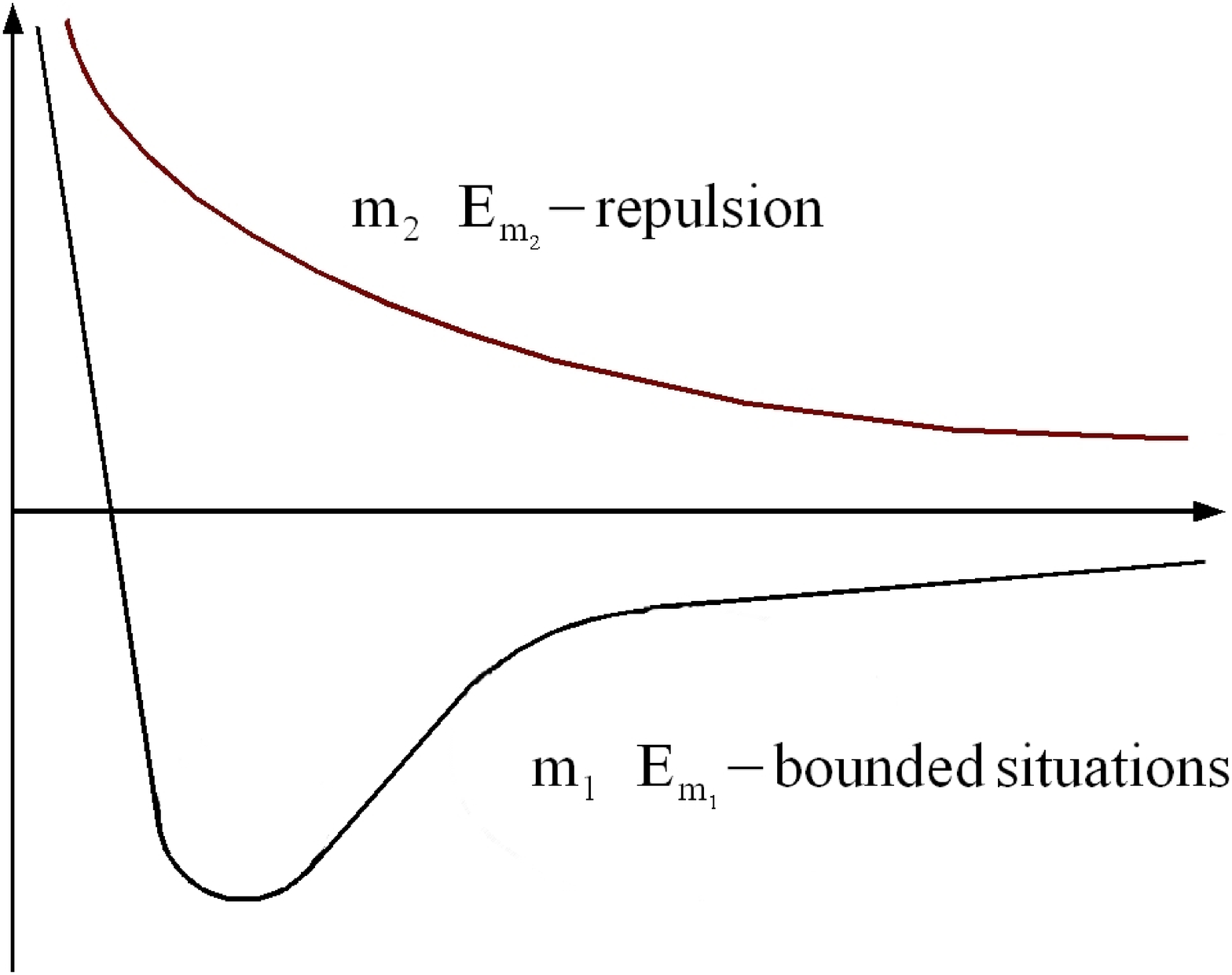}

{\rm Fig. 1}

\includegraphics[scale=0.25]{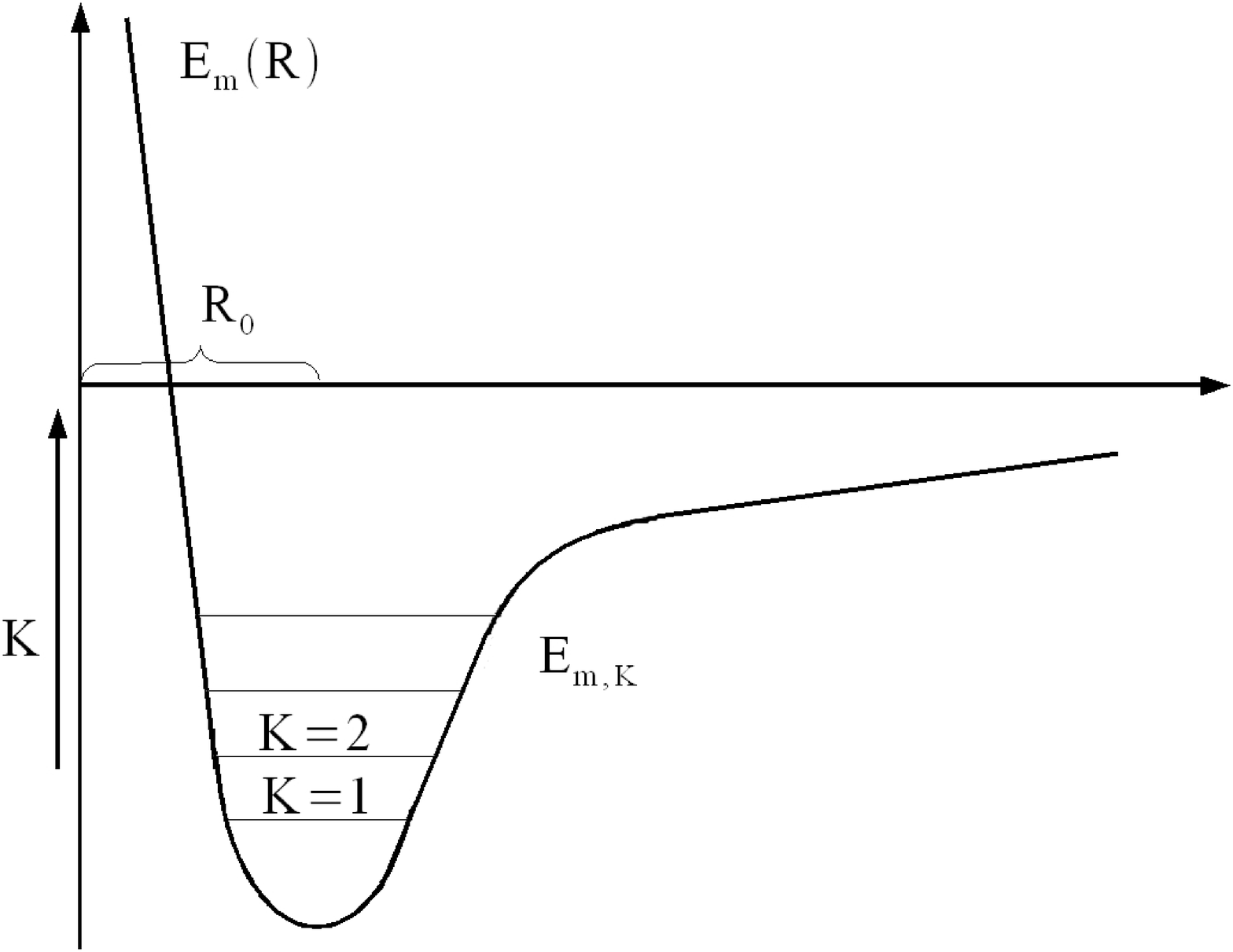}

{\rm Fig. 2}

\includegraphics[scale=0.25]{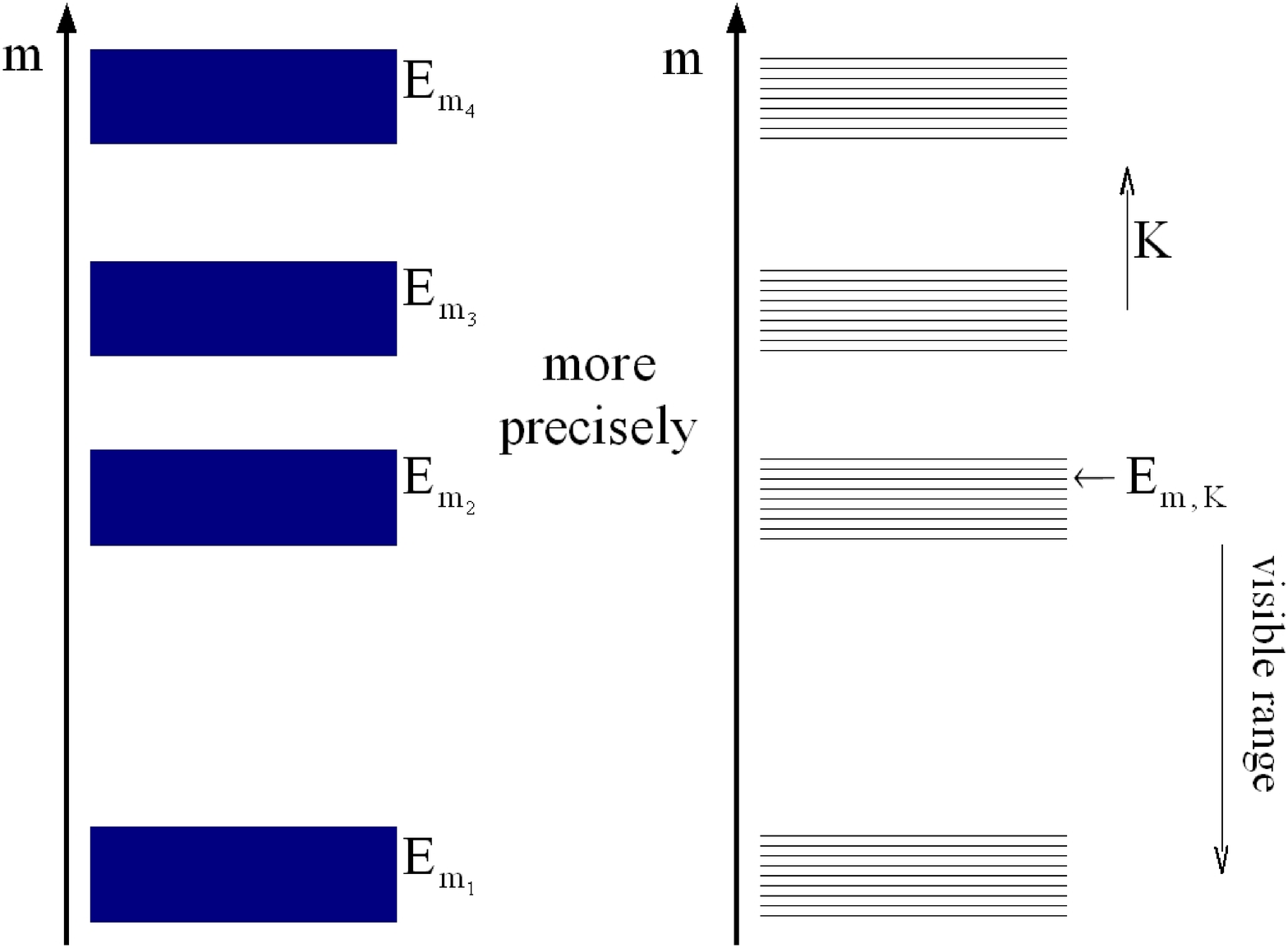}

{\rm Fig. 3}

\includegraphics[scale=0.35]{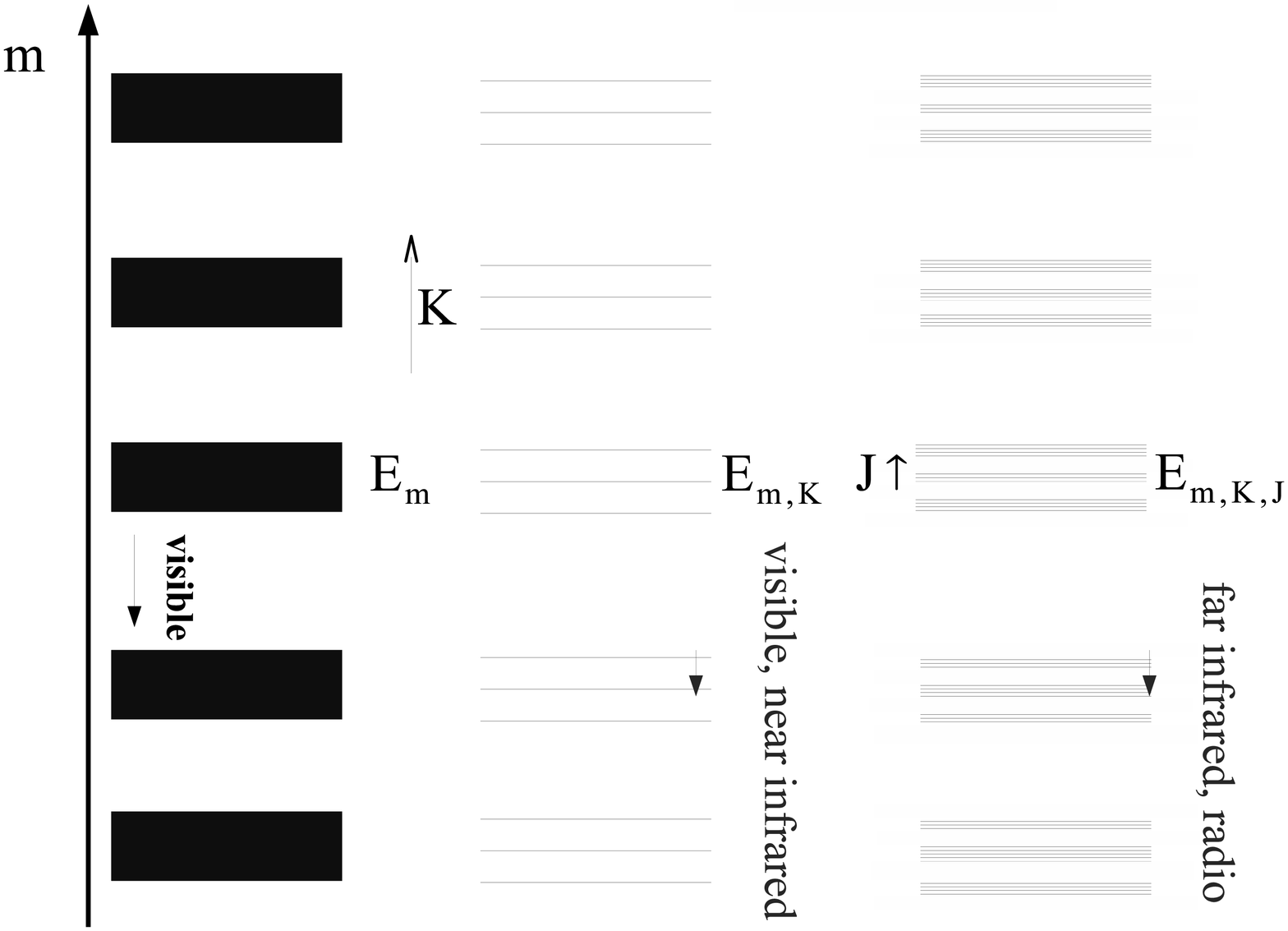}

{\rm Fig. 4}
\end{center}

This situation is often faced with but there are exceptions, when separations between energy levels of various types are comparable. Then one has to do with some resonance phenomena known as the Jahn-Teller effect \cite{Marg_69,Slat_68,Slat_74}. 
\begin{center}
\includegraphics[scale=0.25]{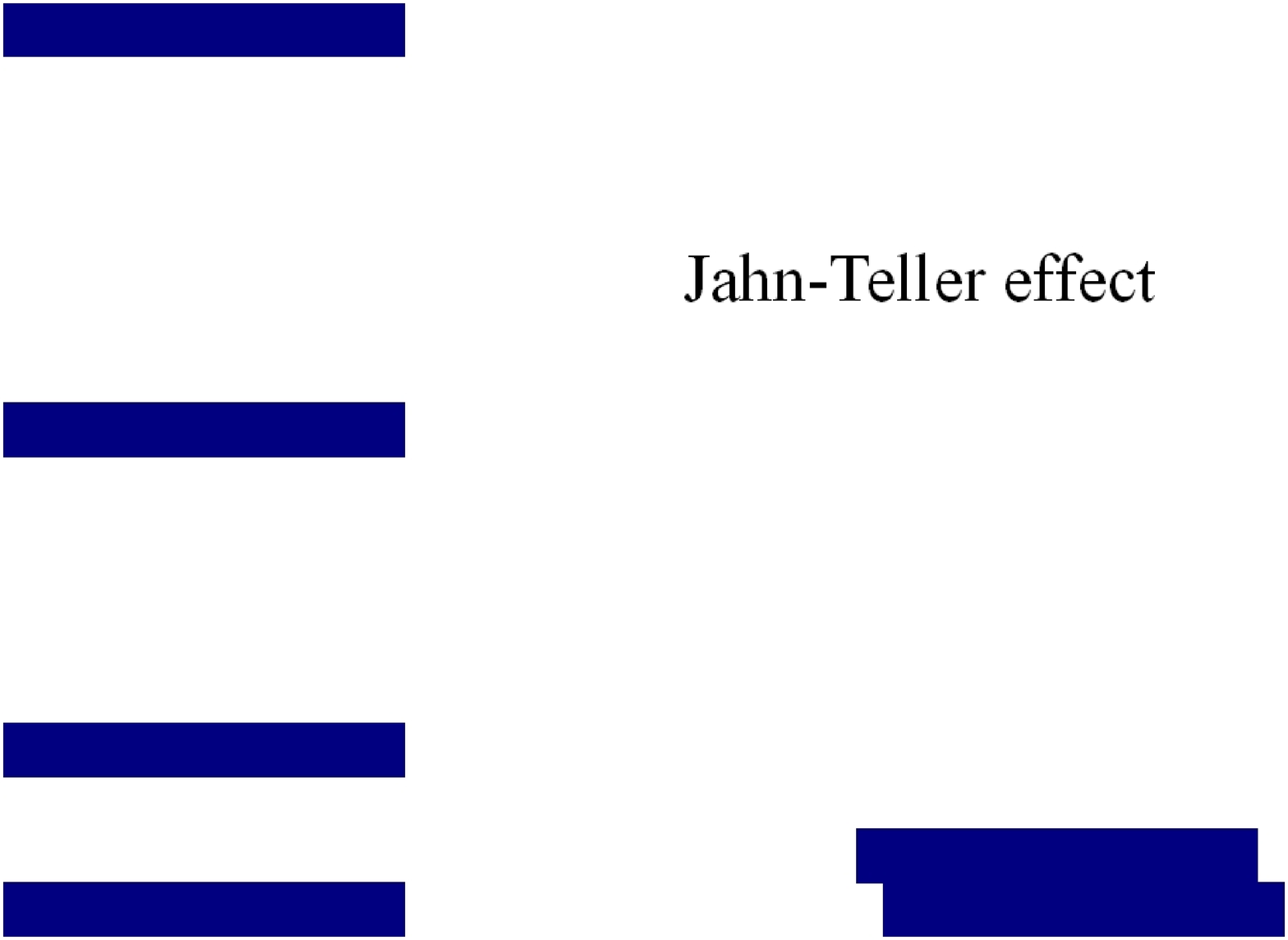}

{\rm Fig. 5}
\end{center}
And even the very Born-Oppenheimer method may be non-applicable in such exceptional situations.

In our model, when applied to molecules, fullerens and similar objects, the general picture and structure of energy levels and their splittings is a bit more complicated and not yet analysed sufficiently in quantitative and qualitative details. The splitting of electronic levels into vibrational and rotational ones has some additional peculiarities. Namely, spatial rotations are controlled by the quantum number $s$, but quantum deformations are described by two things: the spectrum of ${\bf D}_{\lambda}$ and the quantum number $j$ controlling the rotation of squeezing plane of the deformed object.

Moreover, it would be difficult to estimate the structure of splittings in nuclear dynamics, where, nevertheless, some interesting and nontrivial applications are expected.

In our model based on the high affine symmetry one may hope that some partial results may be explicitly obtained. Situation certainly will be much more difficult for more general models of lower dynamical symmetry. The next, more difficult step will follow when even on the level of kinematics we give up the affine model of degrees of freedom and more complicated deformation modes are admitted.

\section{Non-affine modes --- some general and rough comments}

When dealing with molecular dynamics, it is quite natural to expect that affine modes of motion, i.e., rotations and homogeneous deformations, are most relevant for internal phenomena. They are also important in nuclear dynamics. Of course, on both levels the quantization procedure must be carried out.

Nevertheless, any object consisting of more than four material points ($(n+1)$ in $n$-dimensional space) has also other degrees of freedom, even if in a given class of phenomena they are not very important and play a secondary role. In molecular or nuclear dynamics (and also in some macroscopic phenomena) it is often reasonable to establish some hierarchy of degrees of freedom starting from affine ones and then admitting ones more and more complicated. Degrees of freedom of affine motion are represented by the formula (\ref{eq1}) which expresses the Cartesian Euler (current) coordinates as first-order polynomials of Cartesian Lagrange (material) coordinates. It is natural to describe other degrees of freedom in such a way that Euler coordinates are higher-order polynomials of Lagrange variables \cite{D.38}:
\begin{eqnarray}\label{eq0a}
y^{i}\left(t,a\right)&=&{}_{0}q^{i}(t)+{}_{1}q^{i}{}_{A}(t)a^{A}+
{}_{2}q^{i}{}_{AB}(t)a^{A}a^{B}+\cdots\nonumber\\
&=&
\sum^{k}_{p=0}{}_{p}q^{i}{}_{A_{1}\cdots A_{p}}(t)a^{A_{1}}\cdots a^{A_{p}}.
\end{eqnarray}
The coefficients ${}_{p}q^{i}{}_{AB\cdots Z}(t)$ are generalized coordinates; one must remember however that they are symmetric in material (capital) indices, so to be more precise, independent generalized coordinates correspond, e.g., to $A\leq B\leq \cdots\leq Z$. To avoid this redundancy, or rather to reduce it, it is convenient to use the representation in terms of the radial variable and spherical functions \cite{D.38}:
\begin{equation}\label{eq0b}
y^{i}\left(t,a\right)=\sum^{k}_{l=0}
\sum^{l}_{m=-l}q^{i}{}_{lm}(t)|a|^{l}Y^{lm}\left(\frac{a}{|a|}\right),
\end{equation}
where, obviously, $|a|$ is the length of the material radius-vector $a^{K}$ and
\begin{equation}\label{eq0c}
\overline{q^{i}{}_{lm}}=q^{i}{}_{l\: -m},
\end{equation}
because $y^{i}$ are real quantities. Then independent generalized coordinates correspond to $m=0,1,\ldots,l$. If $k>1$, then on the surface of the body more than one deformative waves are formed.

\begin{center}
\includegraphics[scale=0.35]{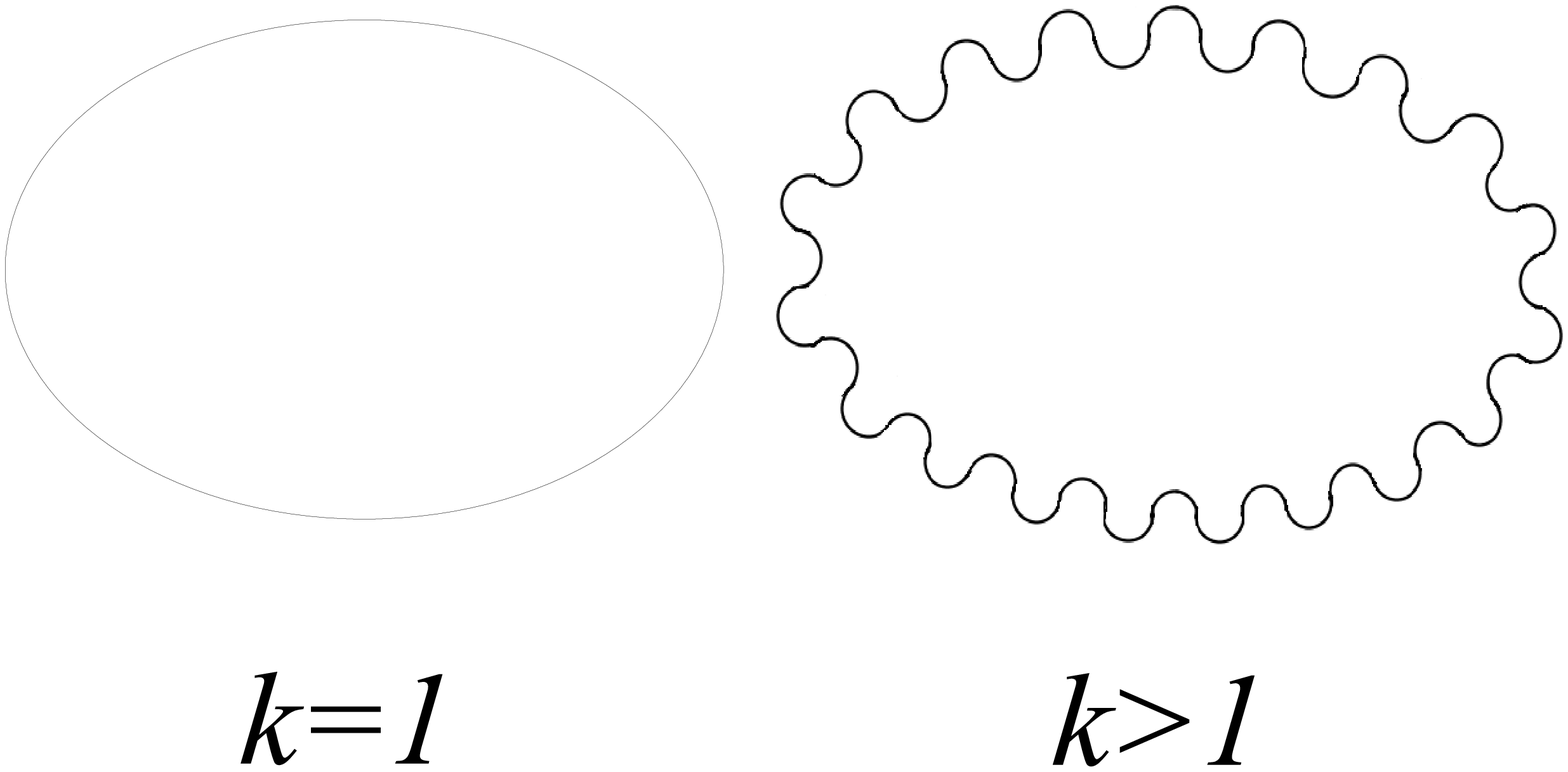}

{\rm Fig. 6}
\end{center}

This procedure is used in so-called method of virial coefficients (widely used in astrophysics, cf. for instance the book by Chandrasekhar). It was also applied in nuclear dynamics, where of course the quantized version of theory must be used \cite{EisGr_70}.

For continuous bodies, $k$ may be in principle arbitrary (even infinite, then $y^{i}$ are simply expressed as analytic functions of Lagrange coordinates). Obviously, for finite systems of material points, $k$ must be finite, because one deals then with a finite number of degrees of freedom. The higher $p$ or $l$ , the less collective character of the corresponding degrees of freedom, although, of course, all of them are "collective" in comparison to individual one-particle positions.

Substituting (\ref{eq0a}) into the formula of the kinetic energy, one expresses it through generalized velocities ${}_{p}\dot{q}^{i}{}_{AB\cdots Z}(t)$ or $\dot{q}^{i}{}_{lm}$; later on Legendre transformation is performed to reformulate everything in phase-space terms, and finally the model is subject to the Schr\"{o}dinger quantization. Obviously, when using polynomial of the order $k>1$ we loose the nice group-theoretical interpretation; it survives only on the level of affine ($k=1$) background phenomena.

{\rm Remark:} if $k>1$, then ${}_{0}q^{i}$ are not coordinates of the centre of mass any longer. Overlooking this fact may lead to serious mistakes.

\section*{Acknowledgements}

This paper partially contains results obtained within the framework of the research project 501 018 32/1992 financed from the Scientific Research Support Fund in 2007-2010. Authors are greatly indebted to the Ministry of Science and Higher Education for this financial support. The support within the framework of Institute internal programme 203 is also greatly acknowledged.

\end{document}